\documentclass[12pt]{iopart}
\usepackage{iopams}
\usepackage{hyperref}
\usepackage[final]{graphicx}
\usepackage{float,color}
\usepackage{here}
\usepackage[normalem]{ulem}
\usepackage{color}

\begin{document}
%\linenumbers

\title{Large-amplitude undulatory swimming near a wall\\
}
\author{Rafael Fern\'andez-Prats$^{1}$, Veronica Raspa$^{2}$, Benjamin Thiria$^{2}$, Francisco Huera-Huarte$^{1}$ and Ramiro Godoy-Diana$^{2}$}
\address{$^1$Department of Mechanical Engineering\\ Universitat Rovira i Virgili, 43007 Tarragona, Spain\\
$^2${Physique et M\'ecanique des Milieux Het\'erog\`enes (PMMH)\\ CNRS UMR 7636; ESPCI ParisTech; UPMC (Paris 6); Univ. Paris Diderot (Paris 7)\\
10 rue Vauquelin, 75231 Paris, Cedex 5, France}}

\ead{rafael.fernandez@urv.cat}

%**************************************************************************************************************
\begin{abstract}
The propulsive dynamics of a flexible undulating foil in a self-propelled swimming configuration near a wall is studied experimentally. Measurements of the swimming speed and the propulsive force are presented, together with image acquisition of the kinematics of the foil and particle image velocimetry (PIV) in its wake. The presence of the wall enhances the cruising velocity in some cases up to 25\% and the thrust by a 45\% {, for swept angles of 160 and 240 degrees}. The physical mechanisms underlying this effect are discussed by studying the vorticity dynamics in the wake of the foil. Proper Orthogonal Decomposition (POD) is applied to the PIV measurements in order to analyse the kinetic energy modal distribution in the flow and to relate it to the propulsion generated by the foil.
\end{abstract}
%**************************************************************************************************************

\submitto{Bioinspiration and Biomimetics}
% Comment out if separate title page not required

\maketitle

%**************************************************************************************************************
%**************************************************************************************************************
\section{Introduction}

Biolocomotion in fluids is in many cases influenced by the presence of a boundary. A well known observation is the case of bird flight near a surface, where the animal can glide with a fixed wing configuration for long distances without loss of altitude \cite{Withers:1977, Blake:1983}. This so-called \emph{ground effect}, which is also of importance in the aerodynamics of aircraft \cite{Staufenbiel:1988} and cars \cite{Katz:2006}, can account in some cases such as the gliding flight of pelicans for induced drag savings of up to $ 50\%$ \cite{Hainsworth:1988}. The physical mechanisms governing the dynamics of the ground effect in such cases where the lifting surface is steady have been extensively studied (see e.g. \cite{Cui:2010} for a short general review or \cite{Rayner:1991} for an in-depth discussion applied to animal flight). The most often cited mechanisms are related to the reduction of downwash in presence of a substrate. In particular,  the fact that induced drag is reduced because wing-tip vortices are inhibited by the presence of the boundary, as well as the enhanced pressure between the lifting surface and the substrate.  Moreover, it has been shown that the ground effect acts to increase not only the lift in steady flight but also the thrust and propulsive efficiency in oscillating modes \cite{Tanida:2001,Quinn:2014b}.

In the case of fish, some species such as batoids swim very close to the substrate, making ground effects an unavoidable element of their locomotor strategy \cite{Blevins:2013}. The main kinematic trait of the pectoral fin of batoids is the production of a backward-propagating wave \cite{Blevins:2012,Rosenberger:2001}, and the physics of the interaction of such an undulating flexible body with a close boundary are likely to be if not completely different, at least significantly modified with respect to their steady counterparts cited above. These issues have only very recently been started to be addressed, for instance using heaving flexible panels \cite{Quinn:2014c} where  the  ground effect was shown to provide notable hydrodynamic benefits in the form of enhanced thrust peaks during the heaving oscillation cycle.  In the same manner as Quinn \emph{et al.} \cite{Quinn:2014c}, the experimental setup used in the present study joins the recent flourishing literature on robotic models using elasticity to mimic fish-like swimming kinematics through a passive mechanism \cite{Alben:2012,Ramananarivo:2013,Dewey:2013,Raspa:2014}.

{In the present manuscript we focus on the effect of swimming near a solid boundary, by studying the self-propulsion of a flexible foil along a rectilinear trajectory actuated by pitching oscillations at the leading-edge. The emphasis is given to the cases with large pitch amplitudes in the head of the foil, that end up developing large deformations in the foil.} Although we focus here in the cruising regimes of our artificial foil, the dynamics of this type of large amplitude undulation influenced by a boundary are certainly a crucial issue for natural or bio-inspired systems on a broader spectrum of swimming regimes, such as the fast-start of fish near a wall \cite{Eaton:1991,Mirjany:2011}. We show that the presence of the wall produces an enhancement of the swimming performance in the large amplitude undulation cases, mainly through a favourable redistribution of momentum in the wake. This effect in terms of cruising velocity can give an enhancement of up to 25\% and defines an optimal position of the foil trajectory parallel to the wall at around 0.4 times the characteristic size of the foil used in the present experiments.

{The main goal of this work is thus to study how the self-propulsion of a model flexible foil performing large amplitude oscillations is affected by the presence of a wall. Experimental measurements of cruise velocities, thrust forces and time-resolved velocity flow fields are analysed. The next section describes the experimental setup and methods and is followed by the presentation and discussion of the results. In addition to performance measurements, based on the trajectory tracking of the foil, Particle Image Velocimetry (PIV) measurements are presented, which permit us to relate the observed effects of swimming near a wall to changes in the wake vortex topology. At the end of the paper we discuss the use of a proper orthogonal decomposition technique to analyse the changes in the energy distribution among the different components of the experimental velocity fields associated to the effect of swimming near the wall.}

%**************************************************************************************************************
%**************************************************************************************************************
\section{Methods}

%**************************************************************************************************************
\subsection{Experimental setup}

\begin{figure}[t]
\centering
\includegraphics[width=0.62\linewidth]{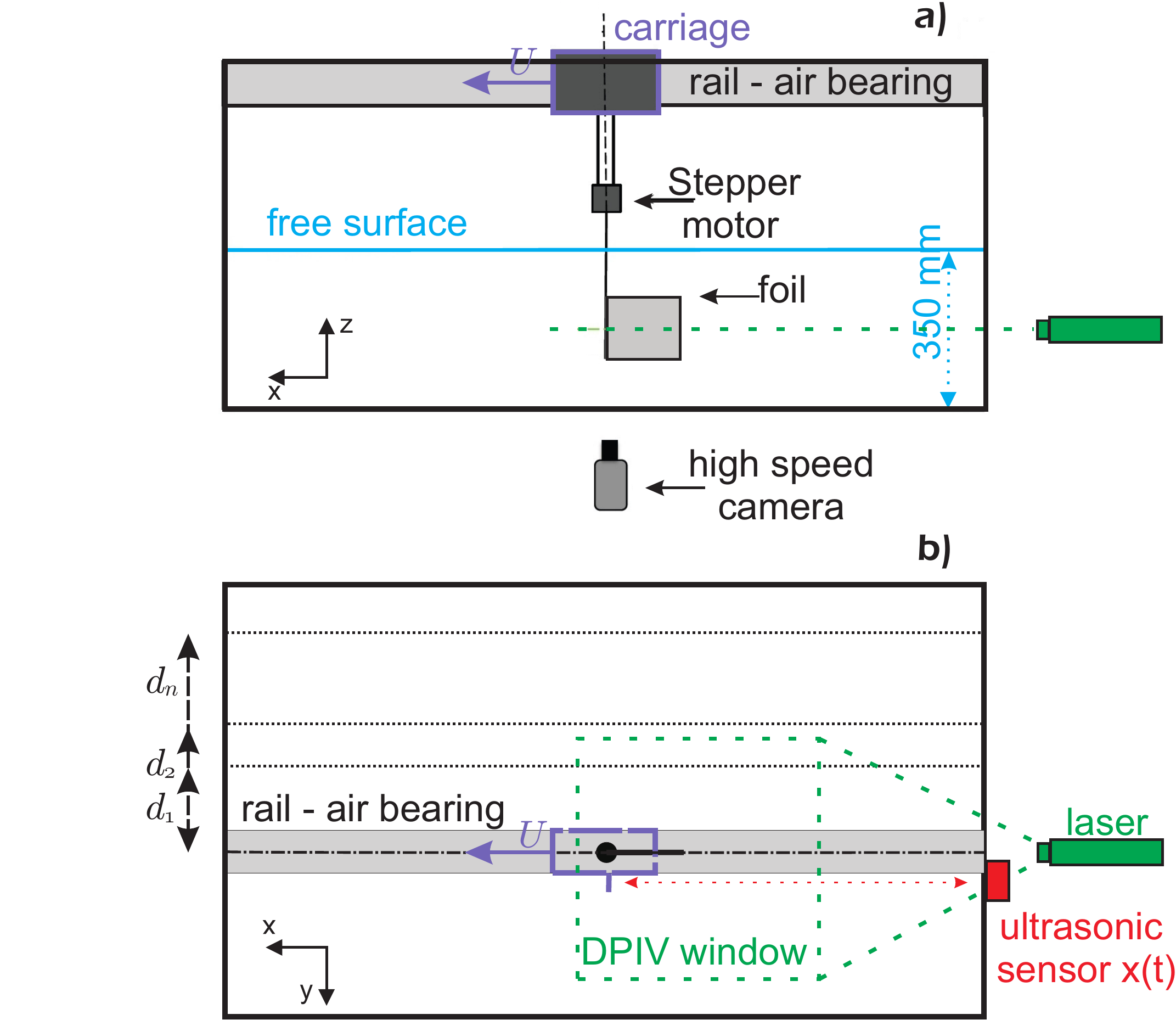} \caption{Experimental set up: (a) lateral and (b) top views.} \label{fig_setup}
\end{figure}

The experiment was conducted in a water tank ($900 \times 800 \times 500$ mm$^3$), where a model of a self-propelled undulatory foil was allowed to move along the rectilinear direction imposed by an air bearing, installed outside the tank (see figure \ref{fig_setup}). The foil was made of a rectangular flexible Mylar foil of thickness 130 $\mu$m, chord $L=110$mm and span $W=100$mm, giving an aspect ratio $AR=W/L=0.9$. The foil was held at one of its edges by a cylindrical shaft of diameter 5mm, acting as the head of the foil. {Although three-dimensional structures are inherent to this type of flows because of edge effects, the quasi-two-dimensional hypothesis can be justified here because of the aspect ratio used for the foil, as other authors have previously suggested \cite{Buchholz:2008}.} The lowest natural frequency of the foil in water $f_0=0.42$ Hz was measured from the response of the foil to an impulse perturbation of the trailing edge as in \cite{Paraz:2014}. A pitching oscillation was imposed through this shaft by means of a stepper motor supported by the moving carriage of the air bearing (see also \cite{Raspa:2013}). A motor driver card was used to control in time the angular position of the shaft, with 0.5$^{\circ}$ of accuracy. A sinusoidal pitch motion was imposed to the shaft yielding to a smooth travelling wave along the foil, providing the desired undulatory kinematics. The self-propelled foil's speed was obtained from time series of the position ($x(t)$), measured using an ultrasonic proximity sensor with an accuracy of 3 mm (see figure \ref{fig_setup}). Additionally, the deformation of the foil was obtained from high-speed video recordings.

The parameters controlled in the experiments were the swept angle ($\theta_0$), the frequency of the pitch motion ($f$) and the gap ($d$). The pitch motion imposed to the shaft or foil's head, can be described by the harmonic expression $\theta=0.5\theta_0\sin{(2\pi f t)}$. {The pitching frequency was stepped with increments of 0.5Hz between each experimental case, except for the case with $\theta$=240 degrees, in which the maximum frequency that the stepper motor could achieve was 3.3Hz instead 3.5Hz.}  The third important experimental parameter was the distance to the wall ($D$), written in dimensionless form ($d=D/L$), using the chord of the foil ($L$). {Six distances to the wall were investigated with dimensionless distances to the wall between 0.55 and 1.54. The strongest effect was observed for separations to the wall in the range 0.25-0.45. The wall effect was considerably weaker for d$>$0.45 with very small velocity and thrust variation. Other distances $d$ were investigated between 0.55 and 1.54 showing practically no differences. Only the largest of those is shown here, corresponding to $d$=1.54.} The Reynolds number based on the foil length ($Re=UL/\nu$), $\nu$ being the kinematic viscosity of the fluid, was between 2200 and 19000. {We recall that the experiment is conducted in a still water tank, so that $U$ is the self-propelled swimming speed and there is no externally imposed free stream, which would have brought the additional effect of the boundary layer near the wall. The latter has been addressed by other authors \cite{Quinn:2014b}, who have studied the effect of the boundary layer in a rigid panel with ground effect.} The parameter space explored for this work ended up in more than 150 experimental cases summarised in Table \ref{t1}.

%Ratio $gap/span$ was chosen according to Webb et al 1993 \cite{Webb:1993}, frequency of the motor and aspect ratio are according to Dewey et al. 2011 \cite{Dewey:2011}.

\begin{table}[t]
\begin{center}
\begin{tabular}{|r|r|r|r|r|r|r|}
\hline
$d$& 0.25 & 0.3 &0.38 &0.45 & 0.55 & 1.54 \\
\hline
$\theta_0$&\multicolumn{6}{|c|}{$f$ (min : step : max)}\\
\hline
40$^{\circ}$ &\multicolumn{6}{c|}{ 1.5 : 0.5 : 5 (Hz) }\\
80$^{\circ}$ &\multicolumn{6}{c|}{ 1 : 0.5 : 4 (Hz) }\\
160$^{\circ}$ &\multicolumn{6}{c|}{ 0.5 : 0.5 : 4 (Hz) }\\
240$^{\circ}$ &\multicolumn{6}{c|}{ 0.5 : 0.5 : 3.3 (Hz) }\\
\hline
\end{tabular}
\end{center}
\caption{Parameters of the experiment.}\label{t1}
\end{table}

\subsection{Particle image velocimetry setup}

In order to investigate the flow around the foil, Digital Particle Image Velocimetry (DPIV) was done to obtain two-dimensional velocity fields. DPIV data were acquired using a system based on a 20mJ Nd-YLF double pulse green laser that produced a planar light sheet, and a high-speed camera at full 1632$\times$1200 pixel resolution, synchronised with the laser in order to capture the illuminated particle cloud images. The flow was seeded using  20 $\mu$ m polyamide particles. A total of 2000 images were recorded for each experiment at a rate of 300, 350 or 400 images/second depending of the frequency of the foil oscillation. Before the velocity fields were calculated, the foil projection was removed from each image by applying a mask able to detect the outline of the foil at each instant in time. Two dimensional velocity fields were computed by applying a Fast Fourier Transform (FFT) based multipass window-deformation technique (\cite{Willert_EiF91}). The algorithm evaluated the images in two steps, first with an interrogation area of $64 \times 64$ pixels and after reducing the size of the window to $40 \times 40$ pixels, all with $50\%$ overlap. Two different types of experiment were measured with DPIV. In some cases, the foil was allowed to move freely along the direction imposed by the rail of the air bearing system (free swimming configuration). In the other type of experiments, the foil was kept at a fixed position by locking the rail of the air bearing system (stationary foil configuration). All DPIV interrogations were made at an horizontal plane located at the middle of the foil's height. {The laser was mounted in the back of the water tank, illuminating the foil from the trailing edge (see Fig.~\ref{fig_setup}b)}. The camera was placed below the tank looking upwards, covering a field of view of approximately 25 cm in the direction of motion and 12.6 cm transversely (see Fig.~\ref{fig_setup}a).

\section{Results and discussion}

%**************************************************************************************************************
\subsection{Foil kinematics}

\begin{figure}[t]
\centering
\includegraphics[width=0.65\linewidth]{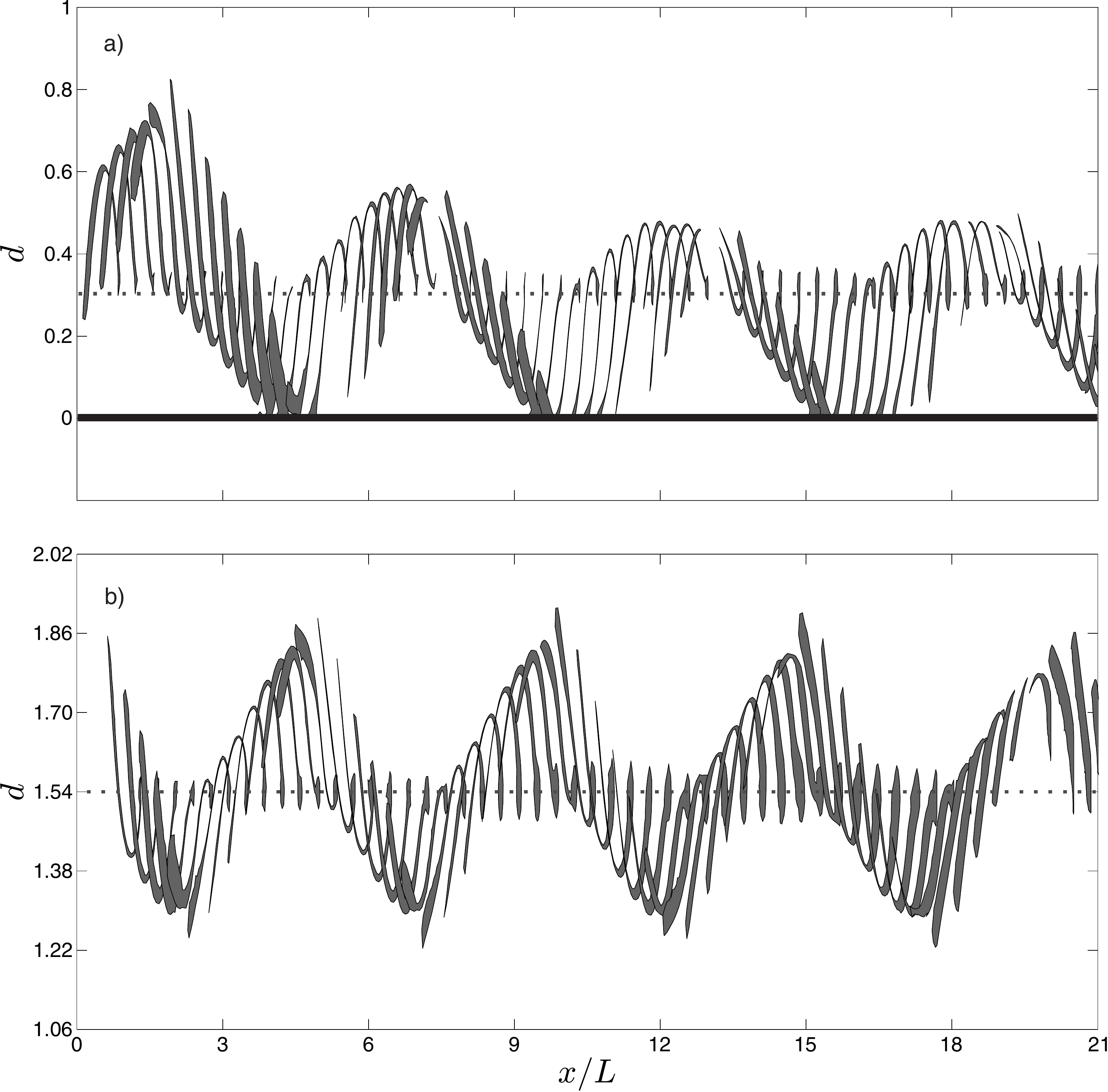} \caption{Sequence of motion of the foil for $\theta_0=160^{\circ}$ and $f=1.5hz$ for two different distances to the wall (a) $d=0.3$ and (b) $1.54$. The foil swims from left to right. The dotted and black lines denote the swimming direction (the trace of the head of the foil) and the position of the wall, respectively.
}\label{kine}
\end{figure}

The undulation, tail amplitude and wavelength are influenced by the distance to the wall and play an important role in the type of the wake and in the swimming performance. In figure \ref{kine} the undulation kinematics of the foil is shown for two experiments: a case near the wall in Fig. ~\ref{kine}(a), and a case with no influence of the wall in Fig.~ \ref{kine}(b). It can be readily seen that the peak-to-peak lateral excursion of the tip is markedly influenced by the presence of the wall.

\begin{figure}
\centering
\includegraphics[width=\linewidth]{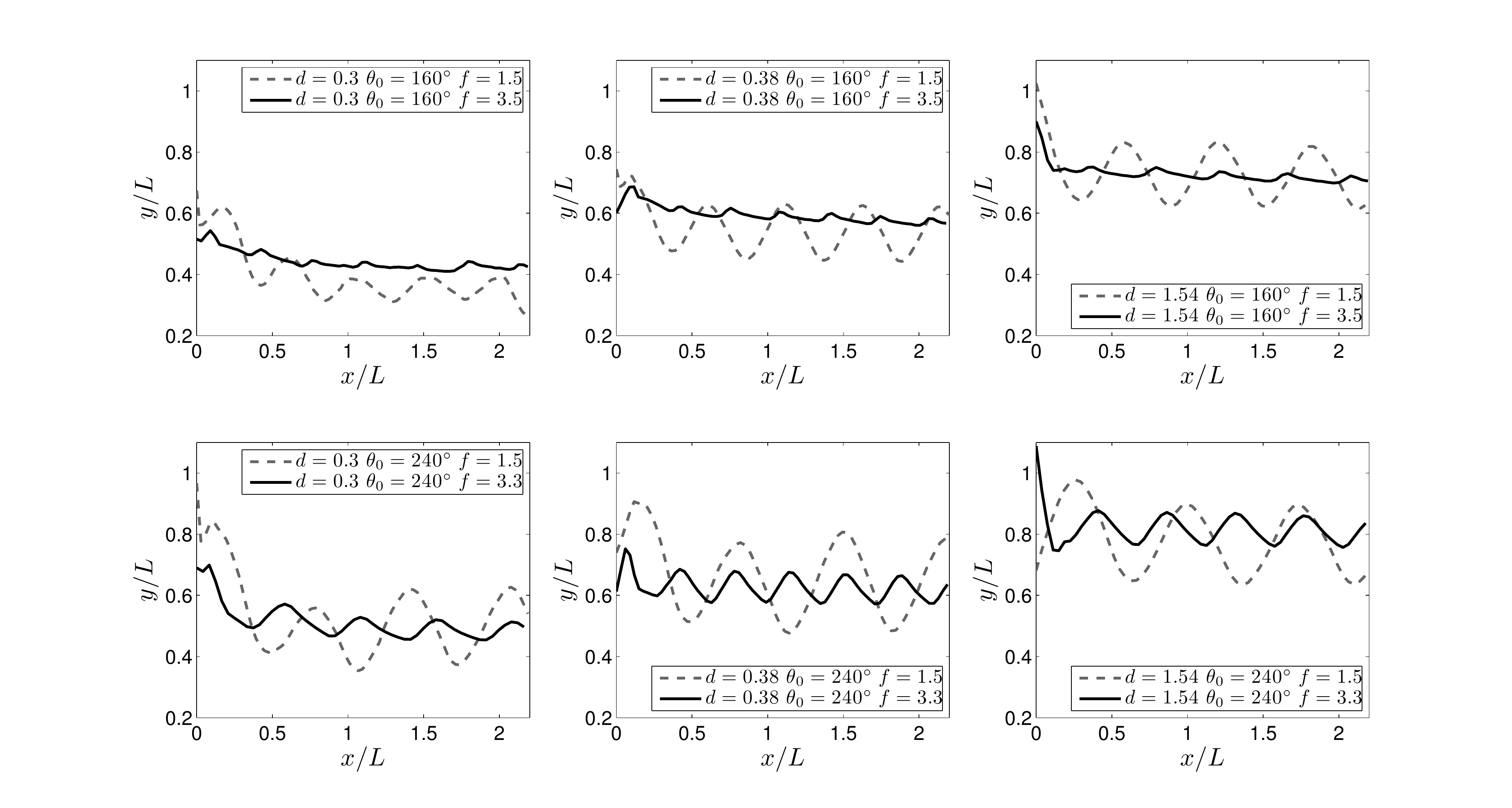} \caption{Envelopes of the trailing edge of the foil motion at different distances to the wall. The parameters for each case are included as a legend in each frame. The top and bottom rows correspond, respectively, to $\theta_0=160^{\circ}$ and $240^{\circ}$. In the left column $d=0.3$, in the centre column $d=0.38$ and in the right column $d=1.54$. Two different frequencies are plotted: $f=1.5$ Hz (dashed line) and 3.3 or 3.5 Hz (solid line).}\label{hl}
\end{figure}

The envelope of the trailing edge of the foil motion is obtained using the Hilbert transform of the time series of figure \ref{kine}. This is shown in figure \ref{hl}, where the top and bottom rows correspond to two different pitch amplitudes of 160$^{\circ}$ and 240$^{\circ}$ respectively, and each column corresponds to one of the three values of $d$ shown previously in Fig.~\ref{kine}.  The two cases with $\theta_0=160$$^{\circ}$ and 240$^{\circ}$ shown are the largest swept angles tested and they mimic the real motion of the backward-propagating wave along an animal  as \cite{Blevins:2013}, \cite{Blevins:2012} and \cite{Rosenberger:2001}. Each graph includes two different pitch frequencies. For both pitch amplitudes, the envelopes show larger amplitudes when the foil is far away from the wall as shown previously by Webb \cite{Webb:1993} and \cite{Webb:2002}. For $\theta_0=240$$^{\circ}$  in figs.~\ref{hl} (d), (e) and (f), envelopes practically do not vary with pitch frequency. On the other hand, for $\theta_0=160$$^{\circ}$,  pitch at a higher frequency produces a smaller envelope amplitude if compared to the low frequency, see Figs.~\ref{hl} (b) and (c). However, this is does not occur close to the wall ---Figs.~\ref{hl} (a)---, here the high frequency generates more amplitude than the low frequency.  The ground effect can be noticed especially at the first peak of the envelope where the amplitude is always higher than the rest of the peaks of cycles as the following graphs a), b), d), and e).

%**************************************************************************************************************

\subsection{Propulsive force and cruise velocity}
\label{S:Forces}

\begin{figure}[t]
\centering
\includegraphics[height=0.41\linewidth]{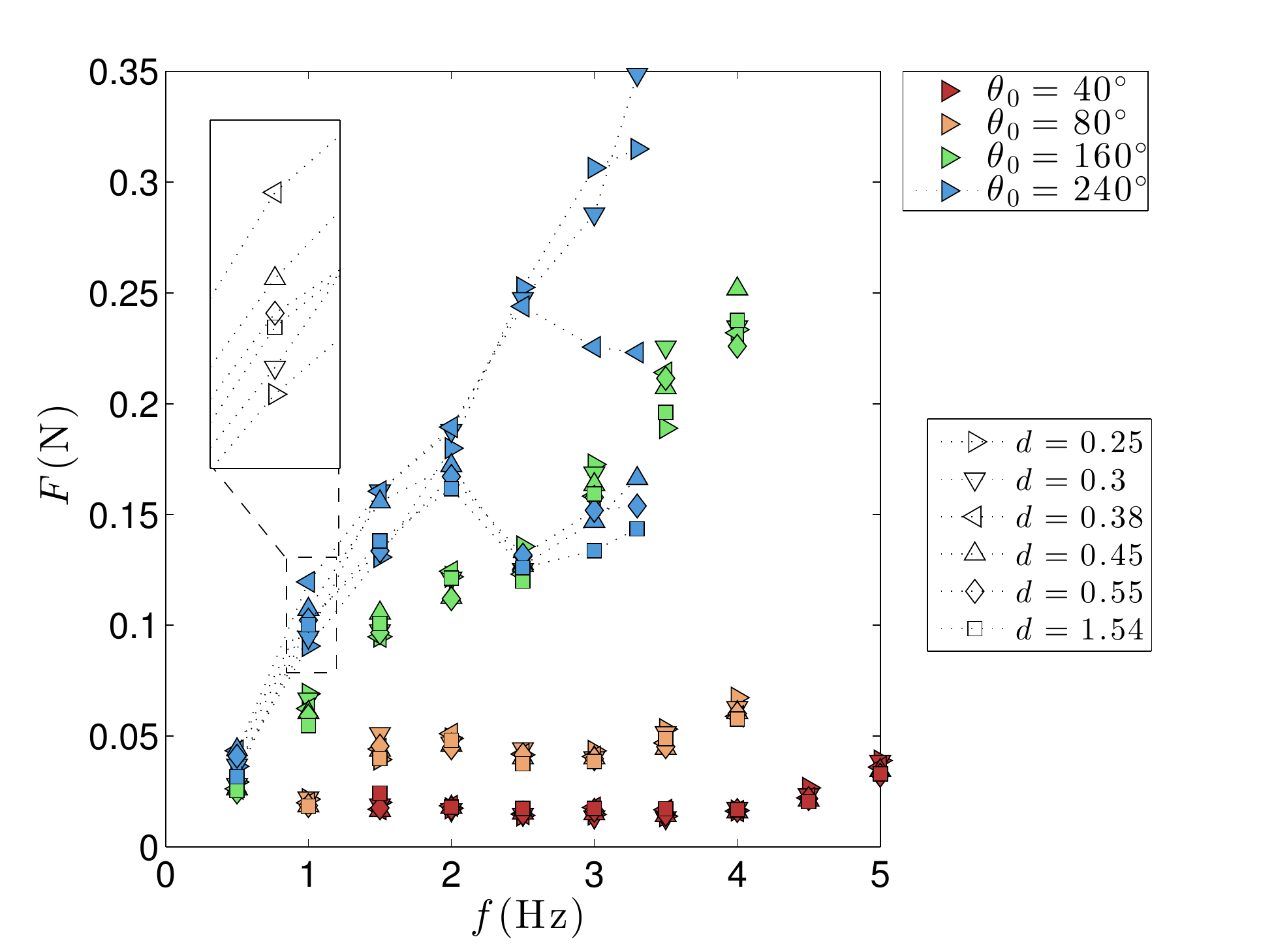}
\includegraphics[height=0.41\linewidth]{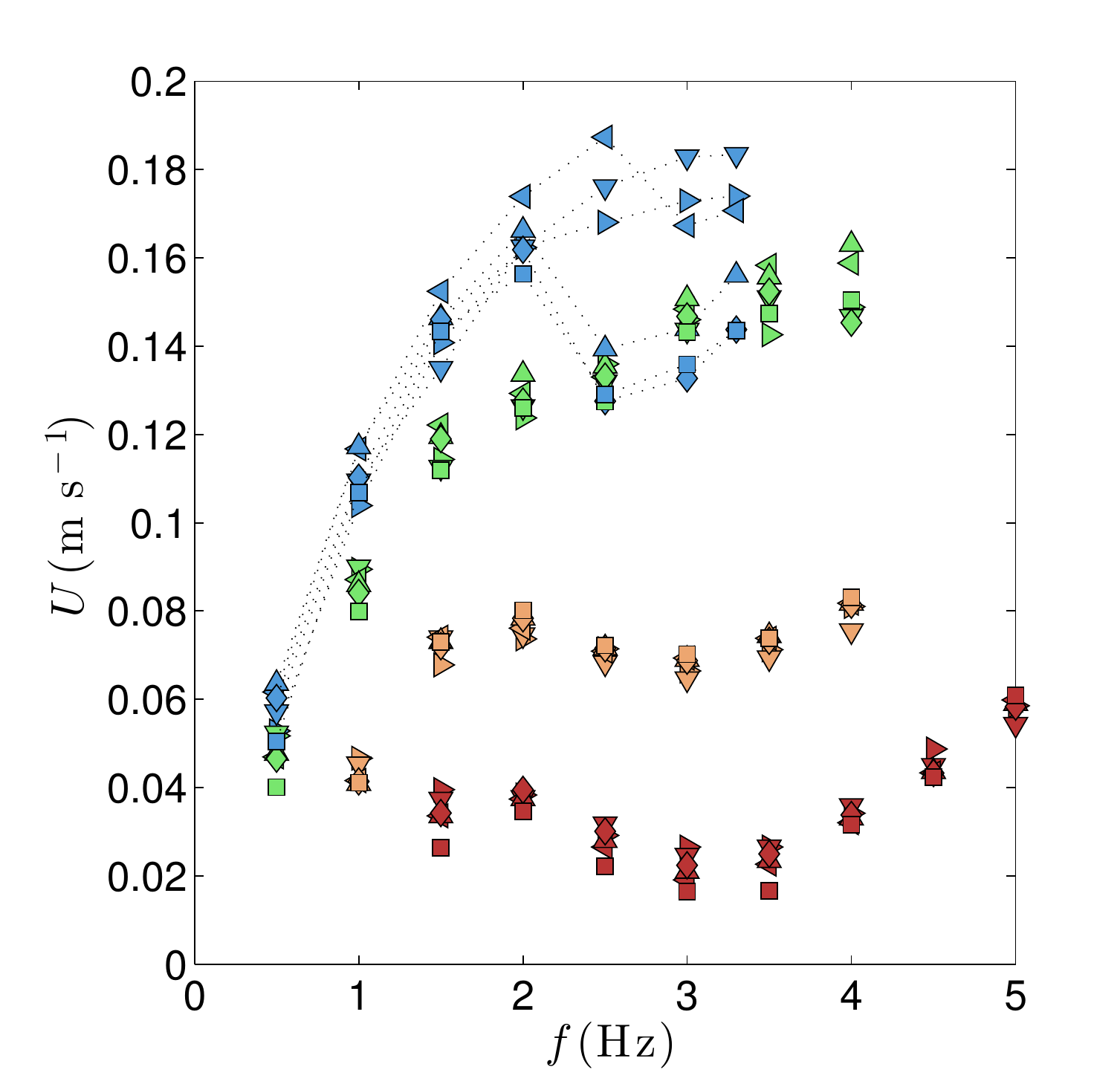}
\caption{(a) Propulsive force (thrust) and  (b) limit velocity (cruise velocity), versus frequency for different swept angles (40, 80, 160, 240 degrees) and distance to the wall (0.25, 0.3, 0.38, 0.45, 0.55 and 1.54). Dotted lines link the data points for each series corresponding to the 240 degrees forcing to guide the eye. } \label{Fpf}
\end{figure}

The propulsive force $F$ and the cruise velocity $U$ are governed by the kinematics of the foil and the distance to the wall. The thrust force $F$ produced by the foil was calculated from the displacement measurements $x(t)$ as in Raspa \emph{et al.} \cite{Raspa:2013,Raspa:2014}. The measured displacement is fitted by the equation $x(t)=\frac{m}{ \gamma}\log \left [ \cosh \frac{\sqrt{\gamma F}}{m} t \right ]$, which is the solution of $m\ddot{x}  + \gamma \dot{x}^2=F$. The latter equation represents a simplified dynamical model of the system in which $m\ddot{x}$ is the inertial term (with a total moving mass $m=2.85$ kg including the body of the foil and its supporting system) and $\gamma \dot{x}^2$ is the hydrodynamic drag term. An iterative optimization process is applied to the analytical solution for $x(t)$, with $\gamma$ and $F$ as unknowns, until estimated and measured values of $x(t)$ converge.

Performance is first analysed by studying how $F$ and $U$ behave as a function of the swept angle $\theta_0$, the imposed pitch frequency $f$ and the dimensionless distance to the wall $d$, see figure \ref{Fpf}. In the figure, different symbols are used to identify distance to the wall, while colours denote the different amplitudes of the pitching oscillation imposed to the head of the foil. The first observation is that the four different sets of pitch amplitudes imposed to the foil, define four distinct branches of performance with respect to frequency. The higher the pitch amplitudes, the higher the swimming speed and the thrust produced. In the two branches corresponding to the smaller pitch amplitudes ($\theta_0=40$$^{\circ}$ and 80$^{\circ}$), the effect of increasing pitch frequency in thrust and cruise velocity is relatively mild, and one recognises the shape of the curves reported in previous studies, with a slight peak that corresponds to a resonant behaviour with one of the deformation modes of the foil \cite{Raspa:2014,Quinn:2014,Paraz:2014}. But when the imposed pitch is large ($\theta_0=160$$^{\circ}$ and especially 240$^{\circ}$), the effect of the forcing frequency is crucial: increasing frequency not only determines more rapid increases in thrust and cruising speed, but also determines that the effect of the proximity to the wall, which was undetectable for the lower amplitudes, appears now as an important element for swimming performance.

\begin{figure}[t]
\centering
\includegraphics[height=0.41\linewidth]{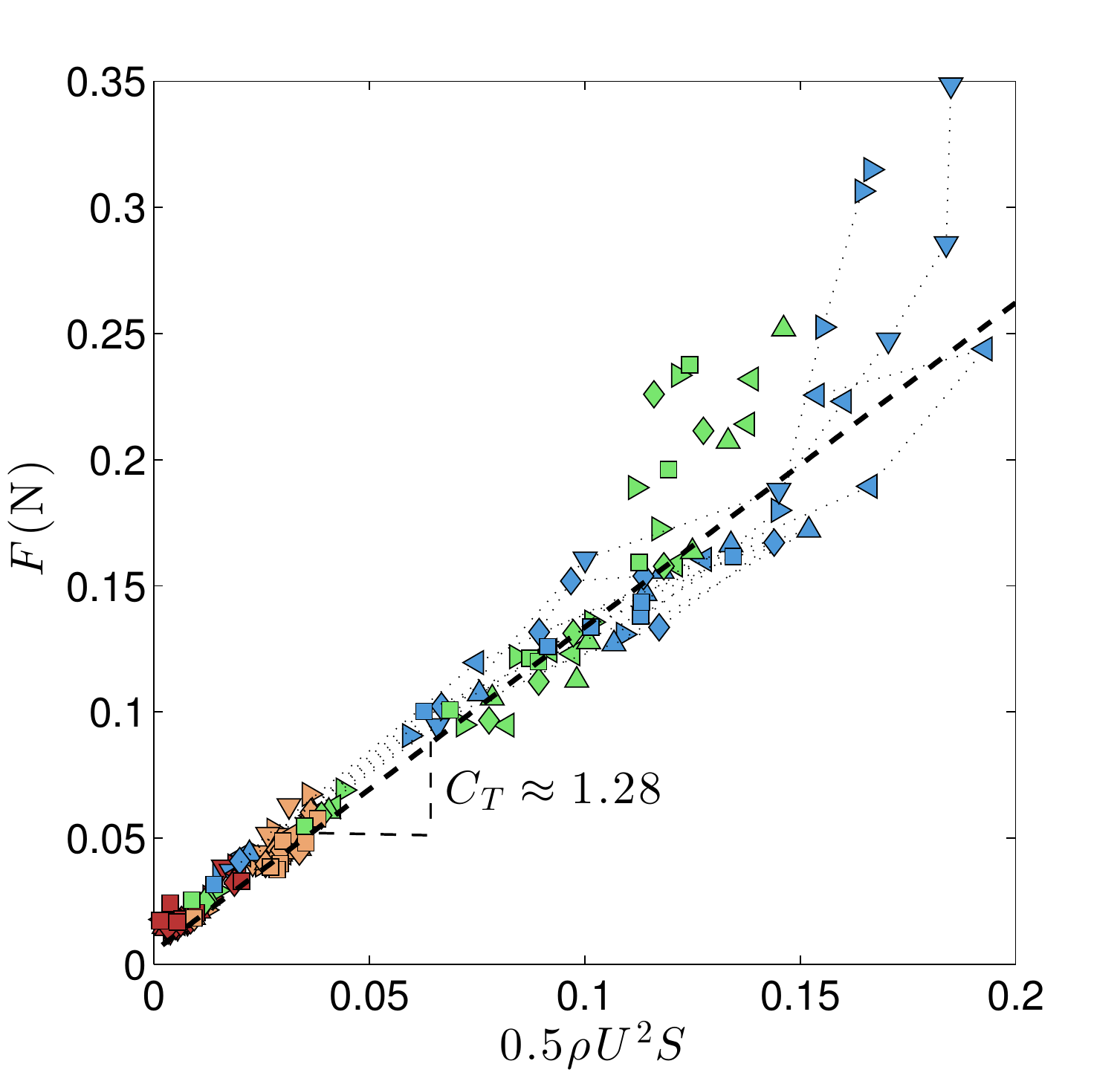}
\includegraphics[height=0.41\linewidth]{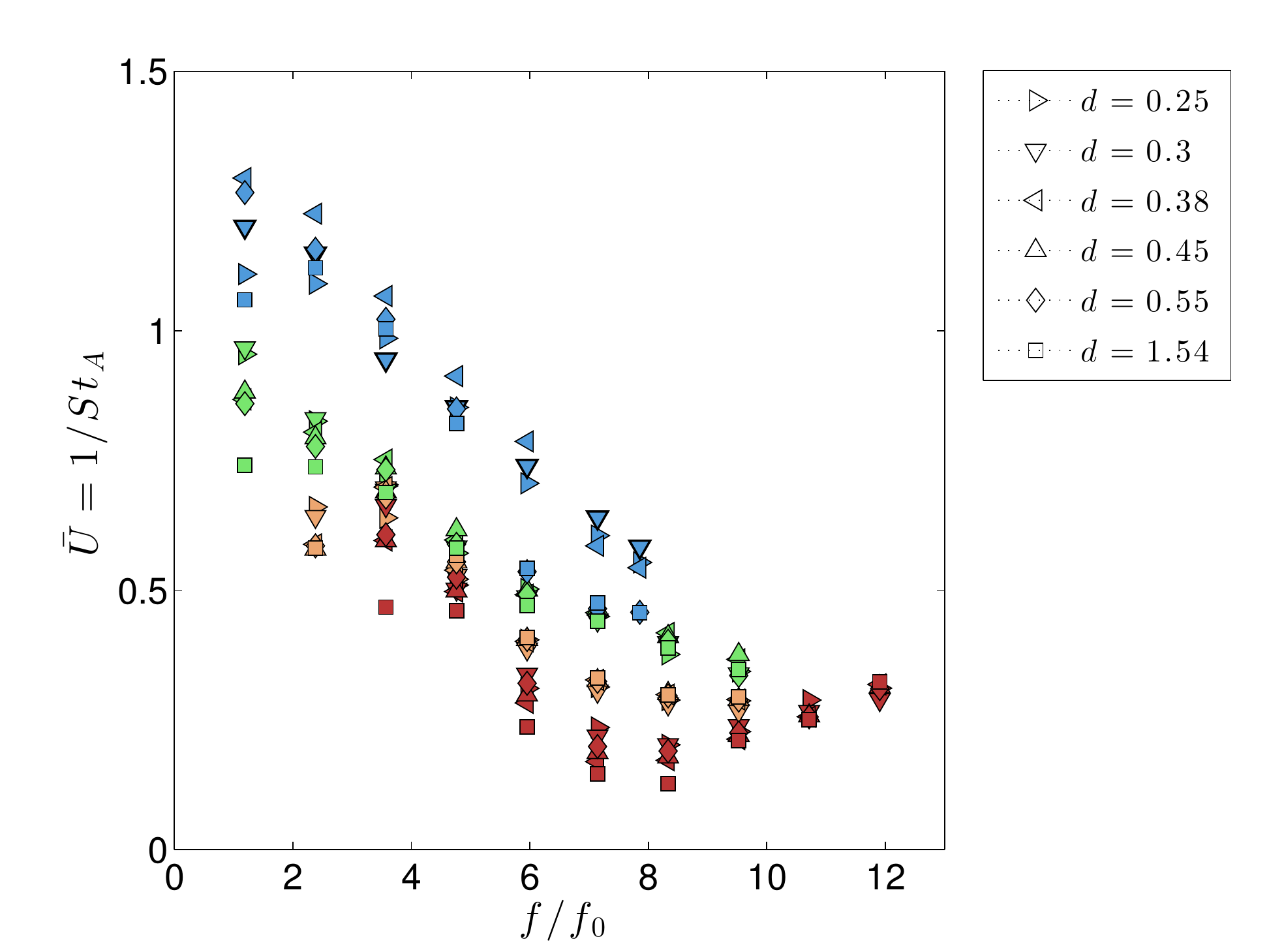}
\caption{(a) $F$ vs. $U^2$ and  (b) reduced velocity $\bar U=U/fA$ as a function of the dimensionless excitation frequency $f/f_0$ for the same data as figure \ref{Fpf} .} \label{DataAdim}
\end{figure}

Considering that the hydrodynamic thrust force at these large Reynolds numbers is expected to scale as the dynamic pressure acting on the propulsive element, the $U$ and $F$ data can be plotted together as $F\propto U^2$ ---see Fig.~\ref{DataAdim}(a)---, where the surface of the foil $S=WL$ and the fluid density $\rho$ have been used in order to obtain a dimensionless thrust coefficient  \begin{equation}
C_T=\frac{2F}{\rho U^2 S} \;.
\end{equation}

The dashed line whose slope is an estimate of the average thrust coefficient was obtained as a linear fit of the data corresponding to $\theta_0=80$$^{\circ}$. It can be seen that while the case of smaller pitching amplitudes ($\theta_0=40$$^{\circ}$) is well described also by this fit, the series corresponding to $\theta_0=160$$^{\circ}$ and $\theta_0=240$$^{\circ}$ deviate notably from the fit roughly for the upper half of the propulsive force range explored in the present experiments. The previous observation is not surprising, since the large amplitude pitching excitation at $\theta_0=160$$^{\circ}$ and 240$^{\circ}$ produces large deformations of the foil, most likely modifying significantly the coefficient of the quadratic drag model used here. Moreover, it is clear from this figure that the proximity to the wall plays thus an important role in the balance of thrust and drag, producing non-trivial behaviours at the large amplitude cases.  Figure \ref{DataAdim}(b) presents another usual way of analysing the self-propelled swimming velocity by means of the reduced velocity $\bar U=U/fA$, a dimensionless parameter   measuring the ratio of swimming speed to a flapping characteristic speed $f\times A$. Here $A=L\sin(\theta_0)$ is the amplitude of the imposed flapping motion.  We note that $\bar U$ is the inverse of the Strouhal number $St_A$ and is related to a ``mechanical efficiency'' of the flapping motion. This representation, however, brings no direct clarification to the role of the proximity to the wall in the scatter of the different data series.

\subsection{Wall effect on swimming velocity}
\label{S:WallEffect}

The effect of the distance to the wall can of course be examined directly comparing the different force or velocity curves in figure \ref{Fpf} as a function of $d$, for each pitching frequency. When the imposed pitch is small ($\theta_0=40$$^{\circ}$ and 80$^{\circ}$), the ground effect is negligible, and all curves collapse over a common curve for each amplitude.  But if the pitch amplitude is increased, swimming near or far away from the wall has a dramatic effect on the thrust and on the cruising velocity. The zoomed region in figure \ref{Fpf}(a) permits to examine as an example the thrust for a foil forced at $\theta_0=240$ and $f=1$ Hz. The maximum value is produced for a distance to the wall $d=0.38$ followed by $d=0.45$, indicating the ground effect is positive. If the distance is too large ($d\geq0.55$) the wall effect starts to be of less importance, becoming negligible at a distance of $d=1.54$, with thrust points collapsing on the same values. This behaviour is in agreement with the observations of \cite{Blevins:2013}. On the other hand, for distances to the wall $d\leq0.3$ the ground effect is negative for thrust. The other notable feature at the largest imposed pitch, $\theta_0=240$$^{\circ}$, is the sudden drop in velocity and thrust when the pitch frequency is set to values larger than 1.5 Hz and the foil is at distances to the wall larger than $d=0.45$. The analysis of velocity fields in the next section will be useful to understand this observation.

\begin{figure}[t]
\centering
\includegraphics[width=\linewidth]{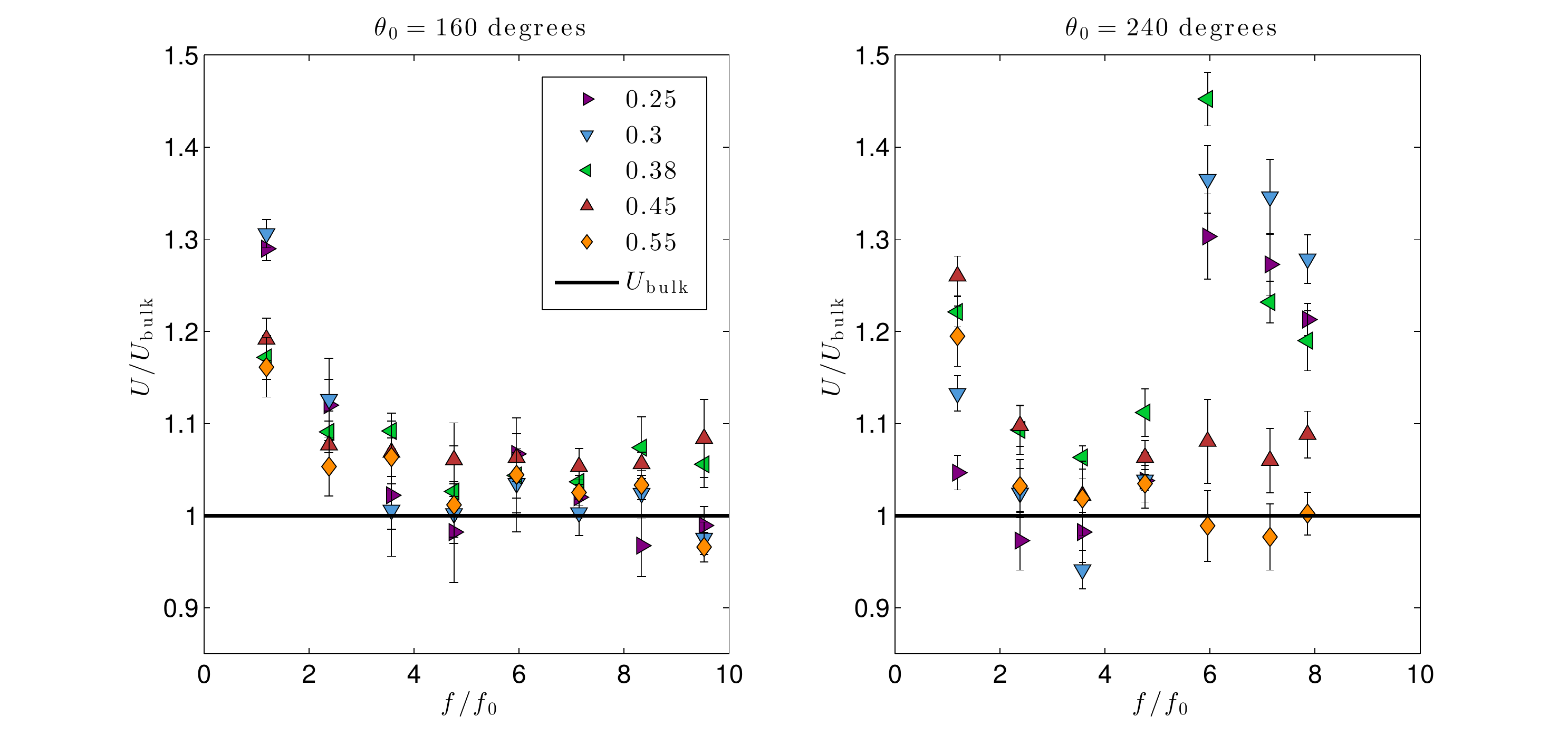}
\includegraphics[width=\linewidth]{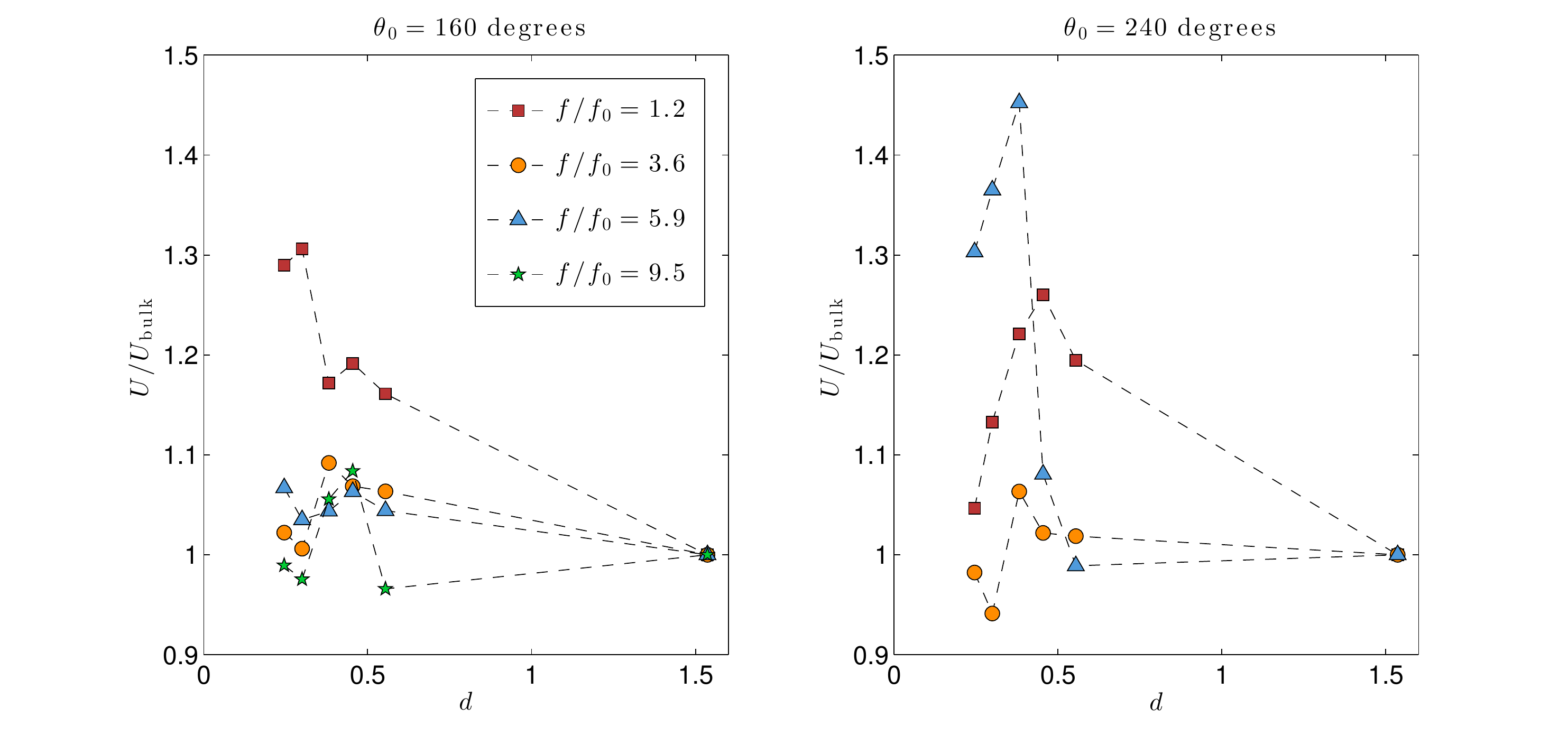}\caption{Cruising swimming velocity rendered dimensionless by normalising it with the cruising speed away from the wall $U_{\mathrm{bulk}}$ as a function of frequency and distance to the wall $d$ (see legends).}\label{vdis}
\end{figure}

Figure \ref{vdis} shows an alternative way of looking at the results, by plotting the cruising speed normalised by its value $U_{\mathrm{bulk}}$ away from the wall (i.e. swimming in the bulk). We focus now on the cases where the effect of the wall is significant which  are those corresponding to $\theta_0=160$$^{\circ}$ and $\theta_0=240$$^{\circ}$. The values of $U/U_{\mathrm{bulk}}$ are plotted against $f/f_0$ for all cases in the the top panels of Fig. \ref{vdis}, the different markers corresponding to different distances to the wall. The two bottom panels of the figure show $U/U_{\mathrm{bulk}}$ as a function of the normalised distance to the wall $d$, only for a few selected frequencies for clarity. Different behaviours are observed for the two different amplitudes analysed and the main features can be summarised as follows: (1) aside from a few exceptions the wall has an overall positive effect on swimming speed; (2) the optimal position with respect to the wall evolves as a function of the frequency and the two different amplitudes tested present different behaviours. For instance, for $\theta_0=160$$^{\circ}$ at the lowest frequency tested, the cases swimming closest to the wall $d=0.25$ -- 0.3 were the best performers, while for $\theta_0=240$$^{\circ}$ the best case was at $d=0.45$; (3) the optimal distance for $\theta_0=240$$^{\circ}$ case presents a sharp change for frequencies higher than $f/f_0\approx 5$, going from $d\approx 0.45$ down to $d\approx 0.3$.

In what follows we examine the velocity field around the swimming foils in order to pinpoint the fluid dynamical mechanisms responsible for the previous observations.

%**************************************************************************************************************
\subsection{DPIV analysis}

DPIV measurements were performed for two different foil configurations: Stationary swimming configuration (air bearing blocked), and self-propelled free swimming configuration (free to swim along the direction prescribed by the supporting air-bearing rail). DPIV was performed for the reference case without wall effect, and for selected cases with wall effect in which there was an enhancement of propulsion, as seen in section \ref{S:Forces}, that is for cases with large pitch motions and moderate distances to the wall. DPIV measurements of the stationary swimming configuration are used to obtain a global overview of the mean velocity fields, whilst in the free swimming configuration, the analysis is focused on the local instantaneous vorticity fields and the different wake topologies found behind the foil. DPIV data in all figures appear in dimensionless form, with velocities given by $(V_{x},V_y)=(v_x,v_y)/fL$ and vorticities computed as $\omega_{z}L/U$.

%**************************************************************************************************************
\subsubsection{Stationary foil}
\label{S:Stationary}

\begin{figure}[t]
\centering
\includegraphics[width=0.7\linewidth]{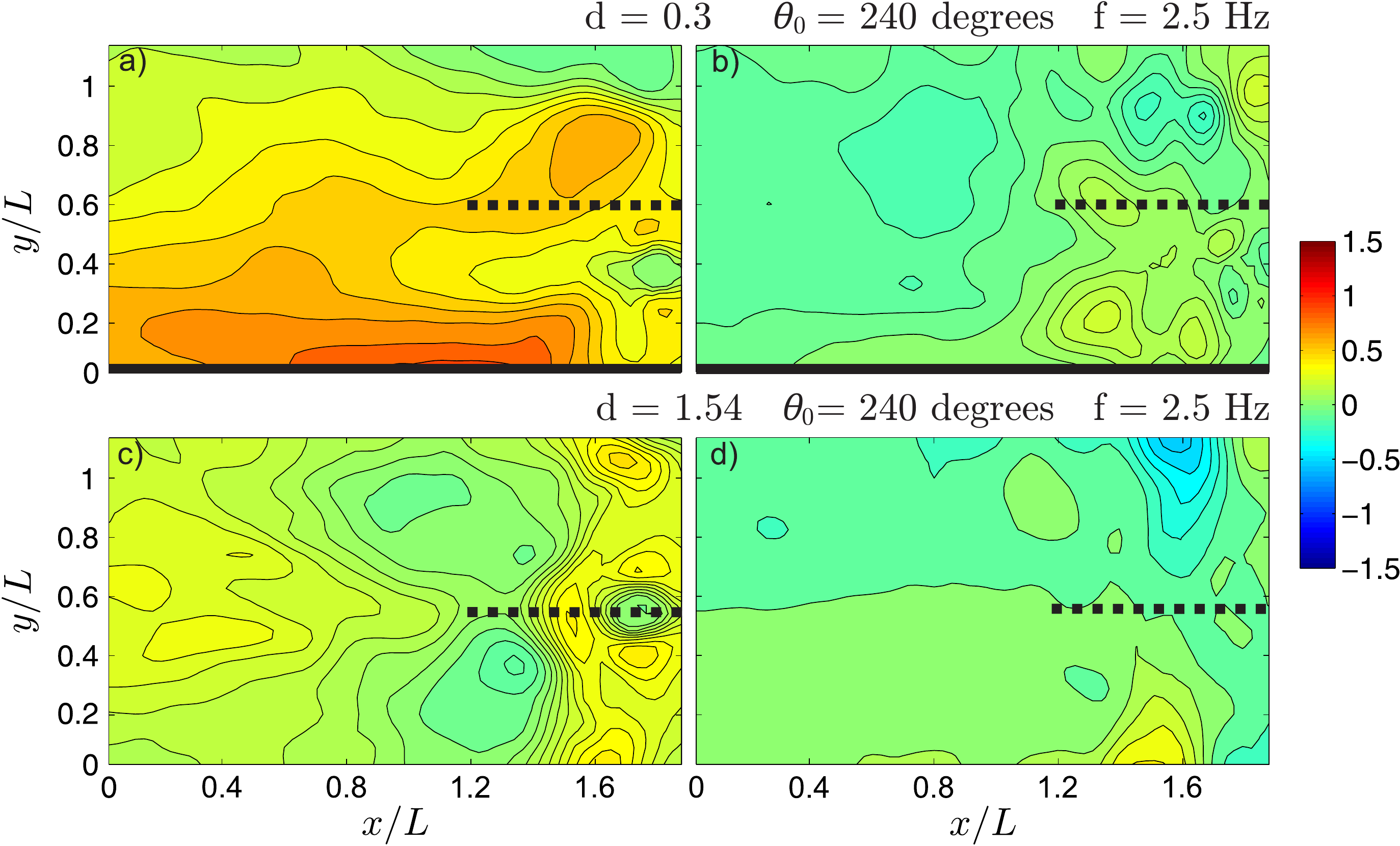} \caption{Average of the velocity fields, stream-wise $\bar V_x$ in the left column and cross-stream $\bar V_y$ on the right column for: (a) and (b) $d=0.3$, $\theta_0=240^{\circ}$ and $f=2.5$ Hz; (c) and (d) $d=1.54$, $\theta_0=240^{\circ}$ and $f=2.5$ Hz. Dashed black lines denote the position of the trailing edge of the foil and black thick lines  represent the wall. The foil swims from left to right.} \label{avgC1}
\end{figure}

\begin{figure}
\centering
\includegraphics[width=0.7\linewidth]{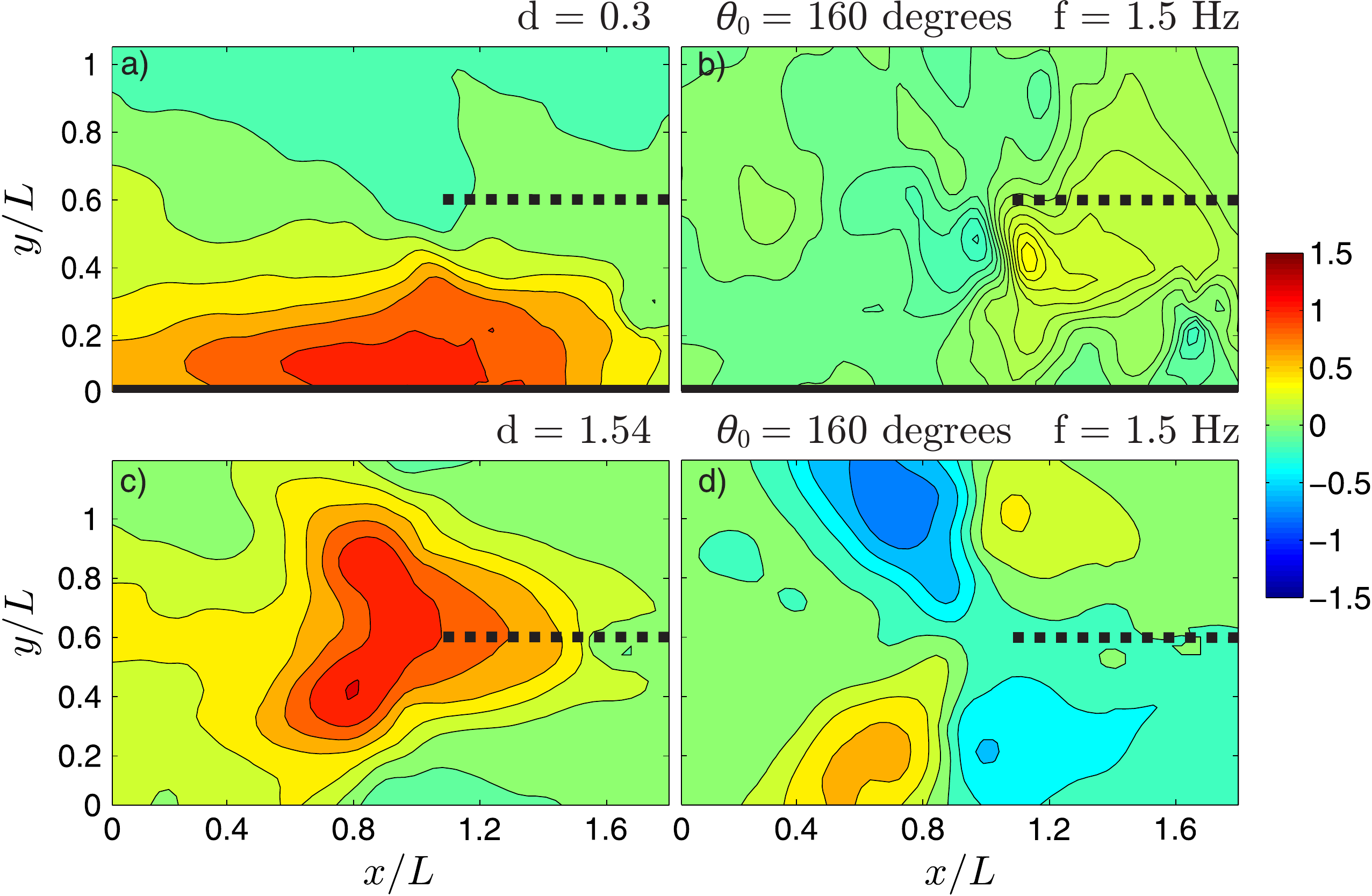} \caption{Average of the velocity fields, stream-wise $\bar V_x$ in the left column and cross-stream $\bar V_y$ on the right column for: (a) and (b) $d=0.3$, $\theta_0=160^{\circ}$ and $f=1.5$ Hz; (c) and (d) $d=1.54$, $\theta_0=160^{\circ}$ and $f=1.5$ Hz. Dashed black lines denote the position of the trailing edge of the foil and black thick lines  represent the wall. The foil swims from left to right.} \label{avgC3}
\end{figure}

Contours of the mean velocity field are presented in figures \ref{avgC1} and \ref{avgC3}, for the stationary foil. The stream-wise component ($V_{x}$) appears in all these figures on the left column and the transverse velocity ($V_{y}$) on the right one. The figures are a good indication of the the momentum distribution in the wake.

Figure \ref{avgC1}  is for an experiment with enhanced propulsion due to the wall effect (plots (a) and (b)) as seen in section \ref{S:Forces} and without wall (plots (c) and (d)) for the $\theta_0=240^{\circ}$ case. The same arrangement of plots appears in figure \ref{avgC3} but for $\theta_0=160^{\circ}$ . There are obvious differences introduced by the wall when comparing by rows the plots in both figures. Whilst in the cases without wall effect in the lower rows, the mean flow fields are typical of those of symmetric wakes \cite{Dewey:2011}, the momentum distribution changes considerably by the effect of the wall, as seen in the upper row of both plots. Regions of high momentum directed along the propulsion direction appear near the wall in both figures, showing clearly one of the causes for propulsion enhancement.

%The changes in the momentum distribution in the wake caused by the wall effect can be seen in more detail in figure \ref{avgC2}. Three experiments carried out with a pitching angle $\theta_0=160^{\circ}$ and $f_0=2.5$ Hz at different distances to the wall are shown. The momentum distribution is clearly greatly modified as a result of the presence of the wall.the foil is placed and changed to the motion direction as the foil becomes closer to the wall. The trailing edge of foil appears depicted in the plots with a black dashed line in its rest position. The last row of plots is for the foil without ground effect and the other two, show the foil as it is placed closed to the wall.
%
%\begin{figure}[H]
%\centering
%\includegraphics[width=0.7\linewidth]{evolution.pdf} \caption{Average of the velocities fields, stream wise in the left column and cross stream at the right column for (a) (b) $d$=0.3, 160 degrees f=2.5$hz$, (c) (d) $d$=0.38, 160 degrees f=2.5$hz$ and (e) (f) $d$=1.54, 160 degrees f=2.5$hz$. Dashed black lines denote the position of the trailing edge of the foil and black thick lines  represent the wall. The foil swims from left to right.} \label{avgC2}
%\end{figure}
%
%**************************************************************************************************************
\subsubsection{Self-propelled free foil}
\label{S:Self-propelled}

In addition to the previous mean-flow analysis with the stationary foil, further insight on the mechanisms that govern the ground effect on swimming performance can be obtained by examining the cases with self-propulsion. In this section the foil is free to move along the rail of the air bearing and DPIV has been used to analyse the instantaneous flow patterns in the wake, depending on the main parameters governing the experiments ($d$, $f$ and $\theta_0$).

\begin{figure}[t]
\centering
(a)\\ \includegraphics[width=0.8\linewidth]{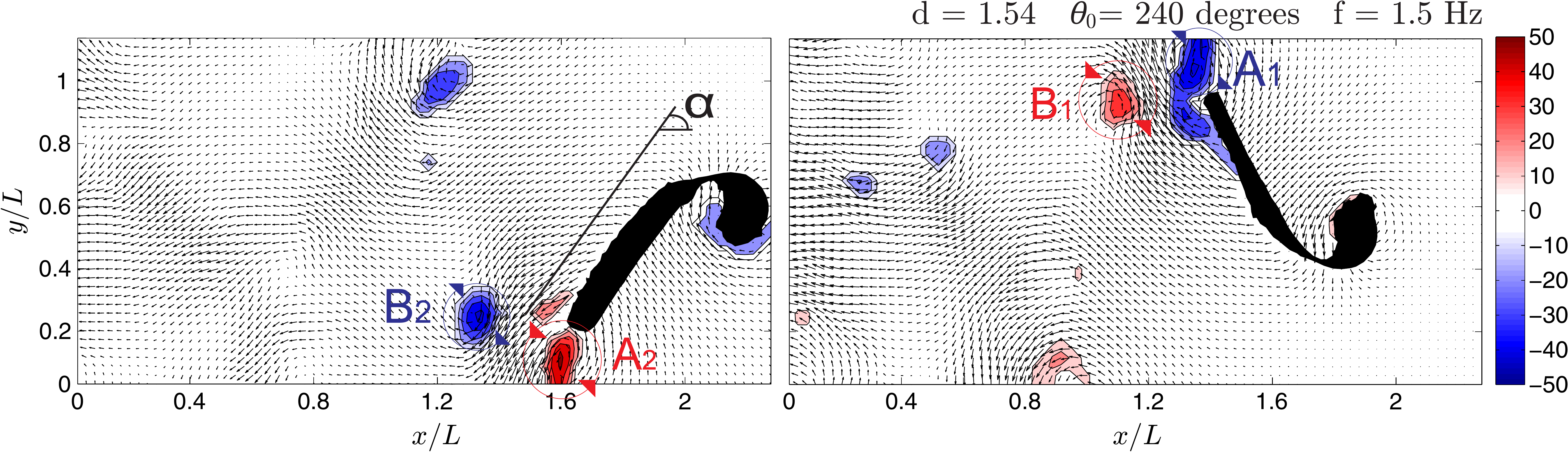}\\
(b)\\ \includegraphics[width=0.8\linewidth]{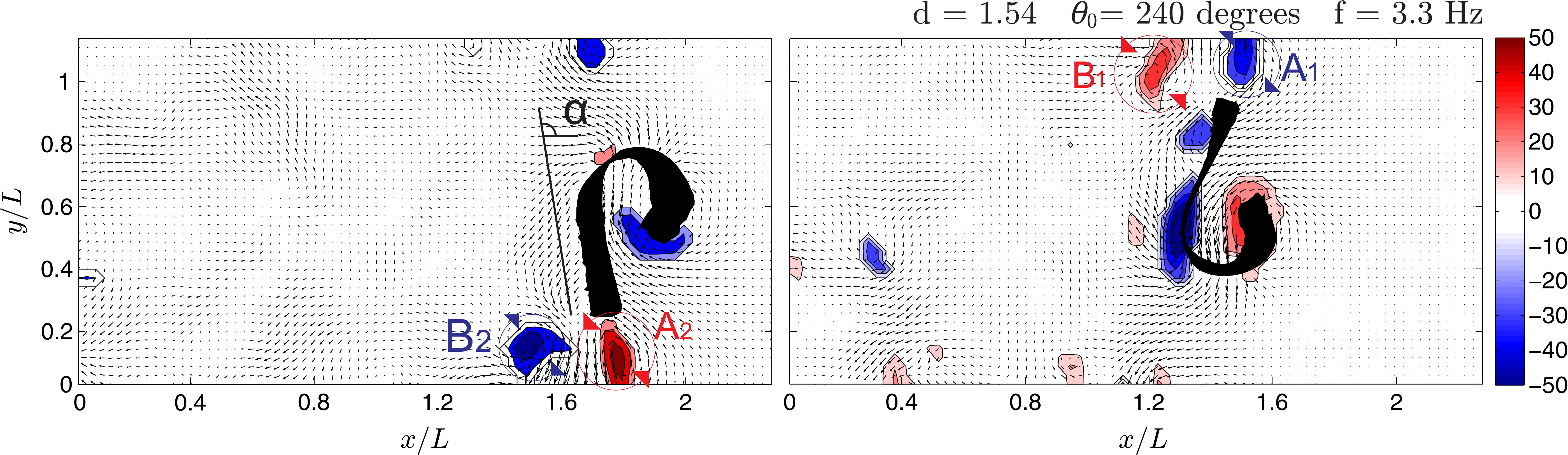}\\
(c)\\ \includegraphics[width=0.8\linewidth]{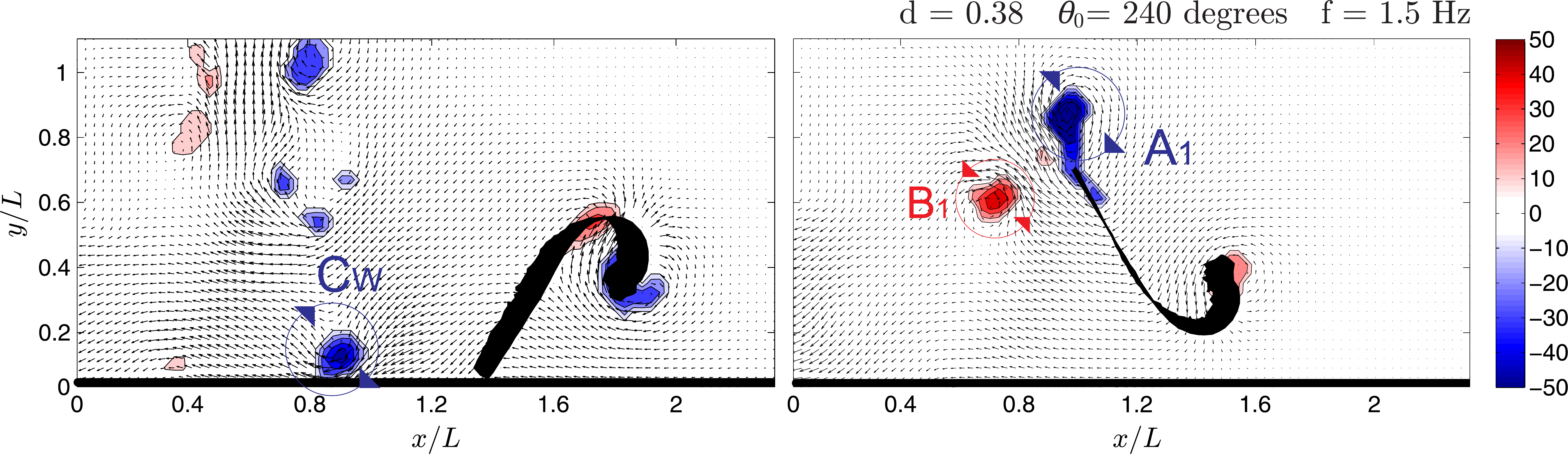}
\caption{Instantaneous vorticity fields and velocity vectors for: (a) $d=1.54$, $\theta_0=240$ degrees and $f= 1.5$Hz; (b) $d=1.54$, $\theta_0=240$ degrees and $f= 3.3$Hz; and (c) $d=0.38$, $\theta_0=240$ degrees and $f= 1.5$Hz. Snapshots at 0$\%$ and 50$\%$ of the cycle are shown on the left and right plots of each row, respectively. The foil swims from left to right. The thick black lines at the bottom in (c) denote the wall. Vorticity colour maps are overlaid on top of the vector velocity field generated by the foil. Blue is used for clockwise vorticity and red is for counter-clockwise. (See text for the description of the vortex labelings in this figure and the following).}
\label{wk1}
\end{figure}

{A nomenclature based on that proposed by Williamson and Roshko \cite{Williamson:1988} to describe the flow structures in the wake of cylinders, is used here to describe the topology of the wake downstream the foil. According to this way of describing wakes, an $S$ is used to denote a single vortex at one side of the wake per shedding cycle. If a $P$ is used, the wake consists of a pair of counter-rotating vortices at one side per shedding cycle. If the same arrangement of vortices is observed at each side of the wake each cycle, a $2$ is placed in front of the $S$ or the $P$. Therefore a $2S$ wake is a wake consisting of a single vortex shed at each side of the wake \footnote{It should be noted that here circulations are reversed with respect to the case of a cylinder wake, the 2S wake being thus the well known reverse B\'enard-von K\'arm\'an (BvK) pattern associated to flapping-based propulsion \cite{Koochesfahani:1987,Anderson:1998,Triantafyllou:2000,GodoyDiana:2008}.} and a $2P$ is a wake made of a pair of counter-rotating vortices at each side. When the observed pattern is different at both sides of the wake, a combination is needed and the symbol $+$ is used. For instance a $P+S$ wake consists of a single vortex in one side and a pair of vortices in the other. In our experiment if there is a combination, the first character before the $+$ symbol denotes the structure observed at the side of the wake without wall, and the second one, after the $+$ indicates the structure at the side of the wall. If the pair of vortices in the $P$ structure is co-rotating, $P^*$ is used.}

\begin{figure}[t]
\centering
(a)\\ \includegraphics[width=0.8\linewidth]{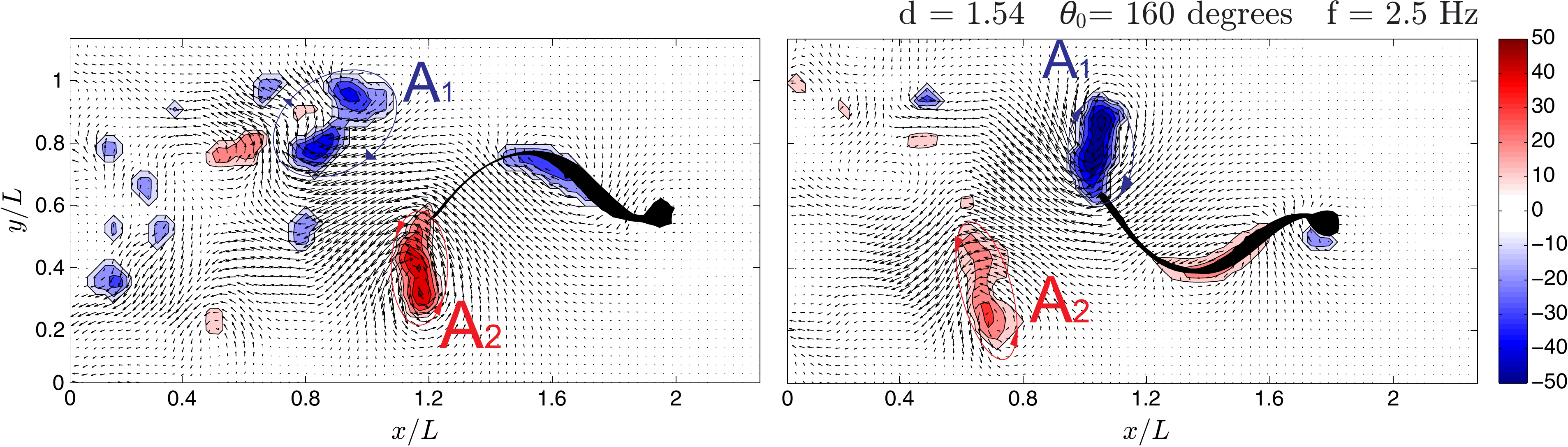}\\
(b)\\ \includegraphics[width=0.8\linewidth]{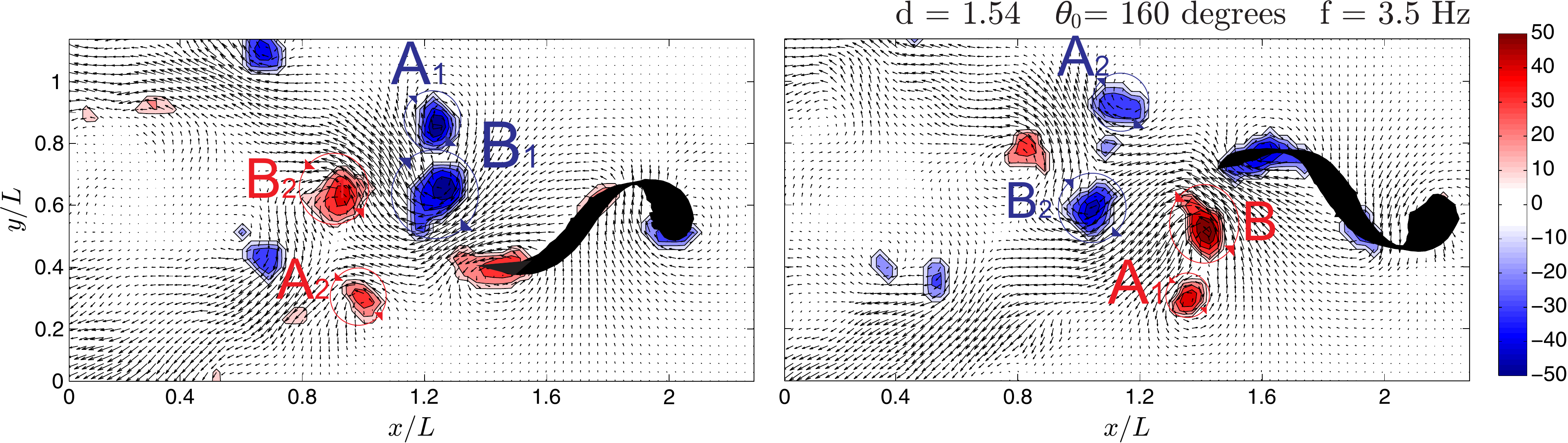}
\caption{Instantaneous vorticity fields and velocity vectors for: (a) $d=1.54$, $\theta_0=160$ degrees and $f= 2.5$Hz; and (b) $d=1.54$, $\theta_0=160$ degrees and $f= 3.5$Hz. Snapshots at 0$\%$ and 50$\%$ of the cycle are shown on the left and right plots of each row, respectively. Other data as in Fig.~\ref{wk1}.}
\label{bvk_wake}
\end{figure}

The patterns observed in the wake of the foil far away from the wall ($d=1.54$, where the wall effect is negligible), are summarised in table \ref{T:bulkVortesModes} for pitch motions of $\theta_0=160^{\circ}$ and $240^{\circ}$ and 3 pitch frequencies. For the case with $\theta_0=240^{\circ}$, the dominant structure in the wake is the $2P$, a pair of counter-rotating vortices at each side of the wake, as observed in figures \ref{wk1}(a) and \ref{wk1}(b) with pitch frequencies of 1.5 and 3.3 Hz respectively. The two vortices in the $2P$ mode are denoted using capital letters and a subscript to indicate each side of the wake, hence $A_1$ and $B_1$ are the vortices at one side of the wake and $A_2$ and $B_2$ are the two vortices at the other side. The figure shows two different instants in time separated half a cycle. For $\theta_0=160^{\circ}$, when the foil is far away from the wall ($d=1.54$), the $2P$ only appears at the lowest frequency ---vorticity field not shown here but similar to figure ~\ref{wk1}(a)---. For higher frequencies $2S$ and $2P^*$ wakes are developed, as shown in figure \ref{bvk_wake}: (a) $2S$ wake with vortices $A_1$ and $A_2$ at each side of the wake, and (b) $2P^*$ wake with two co-rotating vortices at each side of the wake, $A_1$ and $B_1$ at the upper half and $A_2$ and $B_2$ in the lower part.

\begin{table}[t]
\begin{center}
\begin{tabular}{||l | c | r||}
\hline
\hline
d=1.54 & $\theta_0=160$$^{\circ}$  & $\theta_0=240$$^{\circ}$ \\
\hline
\hline
1.5 Hz & $2P$ &  $2P$\\
\hline
2.5 Hz & $2S$ &  $2P$\\
\hline
3.5 Hz & $2P^*$ & $2P$ \\
\hline
\hline
\end{tabular}
\caption{Summary of vortex modes found in the experiments in which the wall effect was not important}
\label{T:bulkVortesModes}
\end{center}
\end{table}

We now describe the vortex wakes observed when the foil is closer to the wall, which are summarised in table \ref{T:WallVortesModes}, focusing on the cases where propulsion was improved: first the case of $\theta_0=240^{\circ}$ and $d=0.38$ and then $\theta_0=160^{\circ}$ and $d=0.3$. In both cases the same pitch frequencies are reported for comparison with the cases presented in table \ref{T:bulkVortesModes} without wall. The patterns are hybrid modes and show complex structures because of the effect of the wall. With the largest pitch amplitude, a $P+S$ structure was observed independently of the pitch frequency. A case showing this $P+S$ structure for $\theta_0=240^{\circ}$, $f=3.5$ Hz and a dimensionless distance to the wall of $d=0.38$ appears in figure \ref{wk1}(c), with vortices $A_1$ and $B_1$ in the upper part of the plot and a single vortex $C_w$ at the side of the wake near the wall. The $2P$ structure observed without wall has now changed to a $P+S$ structure if the wall is near the foil. That is, the counter-rotating vortex pair that was observed at the lower part of the measurement window for the case without wall changes to a single vortex $C_w$ that is pushed vigorously downstream due to the existence of a high momentum jet-like region near the wall. This is readily seen in figure \ref{wk1}(c) by observing the distance at which $C_w$ is located respect to the trailing edge of the foil, compared to the distance of the vortex pair $A_2$ and $B_2$ in figure \ref{wk1}(a).

\begin{table}
\begin{center}
\begin{tabular}{||l | c | r||}
\hline
\hline
Freq & $\theta_0=160$$^{\circ}$ & $\theta_0=240$$^{\circ}$ \\
       & d=0.3 & d=0.38 \\
\hline
\hline
1.5 Hz & $P+S$ &  $P+S$\\
\hline
2.5 Hz & $S+P$  & $P+S$\\
\hline
3.5 Hz & $P^*+S$ & $P+S$ \\
\hline
\hline
\end{tabular}
\caption{Summary of vortex modes found in the experiments in which the wall effect was important.}
\label{T:WallVortesModes}
\end{center}
\end{table}

\begin{figure}
\centering
\includegraphics[width=0.8\linewidth]{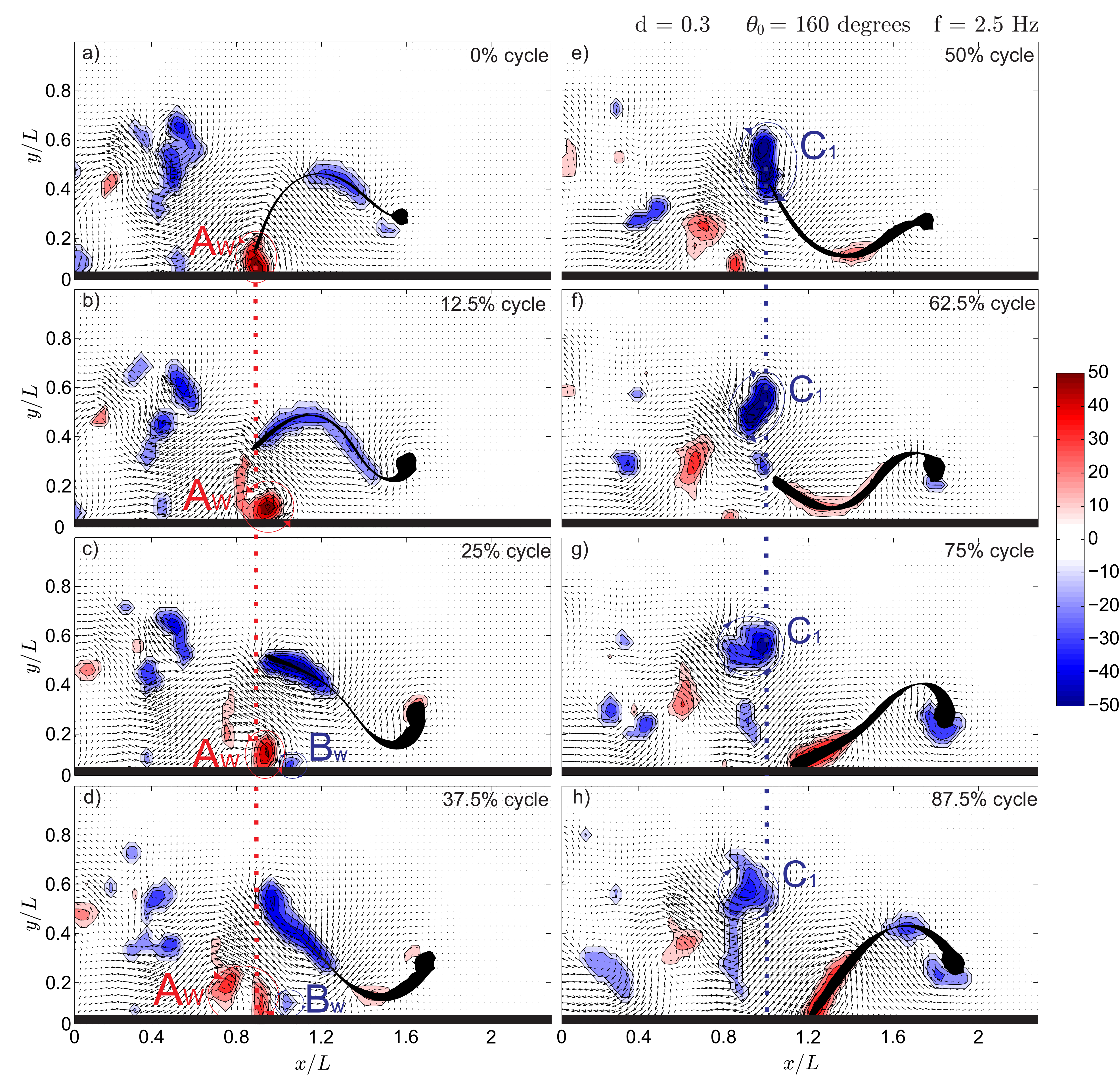} \caption{Sequence of instantaneous vorticity fields and velocity vectors. Every 20 frames is presented ($\Delta$t=50 ms) for $d=0.3$, $\theta_0=160$ degrees and $f= 2.5$Hz. The foil swims from left to right and the black thick lines  represent the wall at $y/L=0$.} \label{wk6}
\end{figure}

\begin{figure}[t]
\centering
\includegraphics[width=0.7\linewidth]{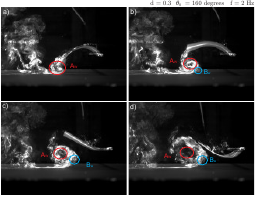} \caption{Flow visualization with fluorescein dye injection and a laser sheet of the vortex structures near the wall effects for $d=0.3$, $\theta_0=160$ degrees and $f= 2$Hz. The time lapse between frames is $\Delta t=250$ ms and the foil moves from left to right.} \label{visu1}
\end{figure}

With the lower amplitude $\theta_0=160^{\circ}$, the structures are clearly dependent on the distance to the wall. At the lowest frequency the $P+S$ is the dominant structure and at a frequency of 2.5 Hz the $S+P$ structure is seen. Figure \ref{wk6} presents a sequence of 8 DPIV snapshots covering a full pitching cycle for the latter case with the foil at a dimensionless distance to the wall of $d=0.3$. At the wall side, a single vortex is shed from the foil ($A_w$) which eventually splits forming another structure $B_w$ because of the proximity to the wall. In the other side of the wake a single vortex $C_1$, forms the $S+P$ mode in the wake. The flow visualisation  with fluorescein-dye presented in figure \ref{visu1} confirms this latter observation and the existence of this counter-rotating vortex pair ($A_w$ and $B_w$) at the side of the wall.

The enhancement in propulsion observed in the thrust and velocity measurements presented above can thus be related to clear changes in the vortex dynamics in the wake of the foil. Whilst at the largest pitch amplitudes the main structure was a $2P$, with ground effect the dominant structure becomes a $P+S$. Now, if the pitch swept angle is 160$^\circ$ the structures are modified to combinations of single and a pair of vortices.

One of the important features observed in the thrust and velocity figures of section \ref{S:Forces}, is the dramatic drop in thrust and foil velocity that takes place at the largest pitch angle as the frequency of pitch is increased. The explanation for that phenomena is clear from figures \ref{wk1}(a) and \ref{wk1}(b) where it can be seen how without the wall, the increase in frequency yields a large change in the angle ($\alpha$ in the figures) at which the shedding of vortices occur, showing that the momentum distribution in the wake becomes less beneficial to the direction of swimming. If the foil is near a wall, the result is a change in this momentum distribution that enhances propulsion: this can be seen comparing \ref{wk1}(a) and \ref{wk1}(c), where without wall, vortices $A_1$ and $B_1$ remain unchanged, but with the wall the disappearance of $A_2$ and $B_2$ to form $C_w$, indicates less energy is dissipated in the wake and a higher-momentum jet-like structure is produced near the wall. This more beneficial momentum distribution was also pointed out in the analysis of the averaged flow fields presented for the stationary configuration in section \ref{S:Stationary}.

%**************************************************************************************************************
\subsection{SPOD analysis}
\begin{figure}[t]
\centering
\includegraphics[width=0.7\linewidth]{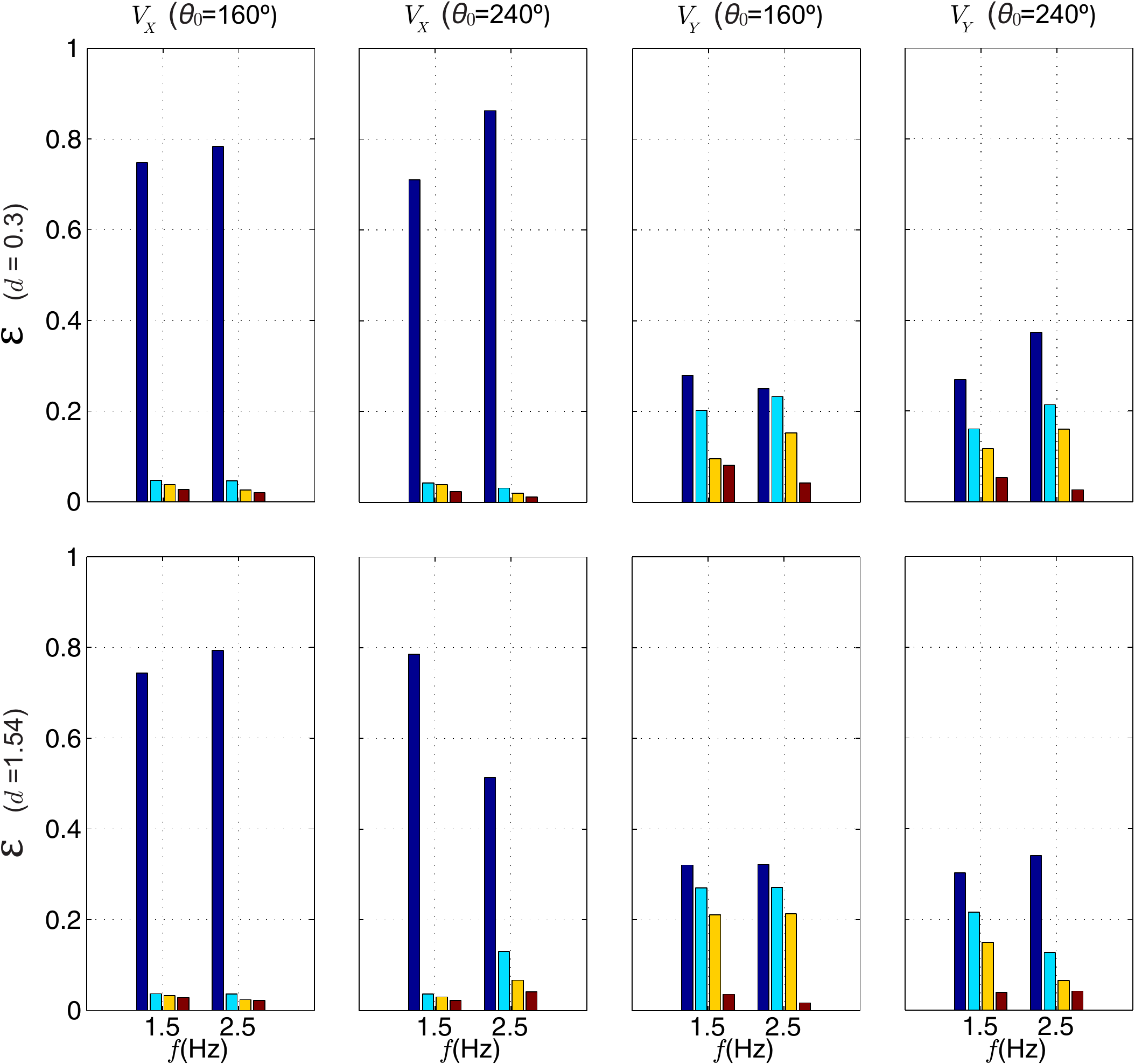} \caption{POD kinetic energy $V_{x}$ and $V_{y}$ of the first four modes (ordered from high to less energy from left to right on each plot) versus frequency (1.5 and 2.5 Hz). First row for $d=0.3$ and second row for $d=1.54$. First and second columns stream-wise velocity and the third and four columns for cross-stream velocity.}
\label{PODt}
\end{figure}

\begin{figure}[t]
\centering
\includegraphics[width=0.63\linewidth]{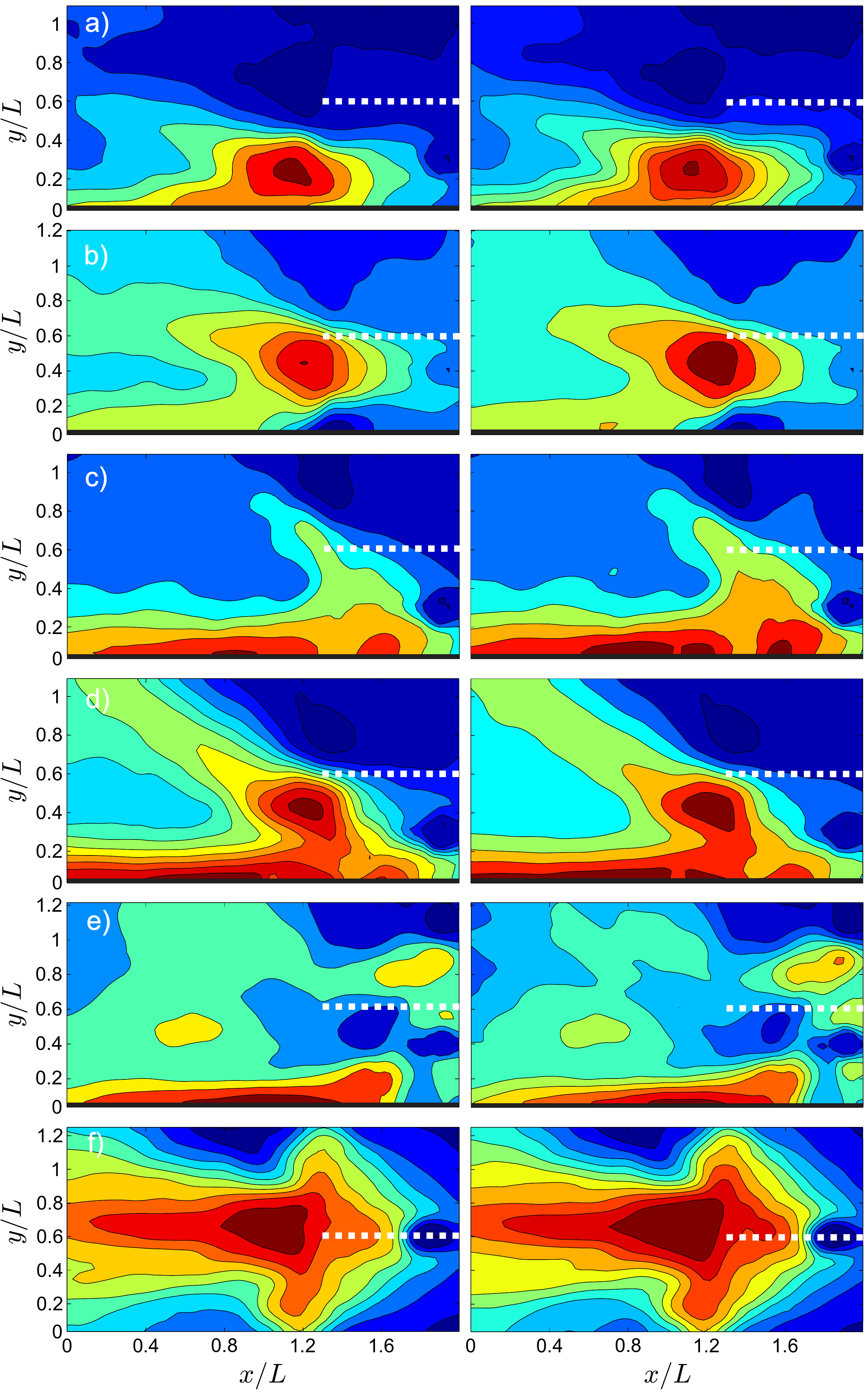} \caption{ Comparison first POD mode (left column) and average of velocity fields (right column) for the stream-wise direction. a) $d=0.3$, $\theta_0=160$ degrees and $f= 2$Hz b) $d=0.38$, $\theta_0=160$ degrees and $f= 3$Hz c) $d=0.3$, $\theta_0=240$ degrees and $f= 1.5$Hz d) $d=0.3$, $\theta_0=240$ degrees and $f= 2.5$Hz     e) $d=0.38$, $\theta_0=240$ degrees and $f= 3$Hz f) $d=1.54$, $\theta_0=240$ degrees and $f= 1.5$Hz. Dashed white lines denote the position of the trailing edge of the foil and black thick lines  represent the wall. The foil swims from left to right.} \label{PD}
\end{figure}

Snapshot Proper Orthogonal Decomposition (SPOD) \cite{Sirovich:1987} has been applied to the velocity DPIV data, following the technique described by Huera-Huarte et al. \cite{HueraHuarte:2009} and recalled in Appendix I.  Assuming that the fluctuating part of the flow can be represented by linear combinations of POD modes $\phi_i(x,y)$ and time varying modal coefficients $a_i(t)$,

\begin{equation}
V(x,y,t)=\bar{V}(x,y)+\sum_{i=1}^M a_i(t) \phi_i(x,y)
\label{E:BasisPOD}
\end{equation}

\noindent the SPOD technique permits to study the kinetic energy ($\varepsilon$) distribution of the flow into the most important modes. The $\varepsilon$ associated to the first four more energetic POD modes is shown in figure \ref{PODt}, for both stream-wise and cross-flow components of the velocity. Two different dimensionless distances to the wall (0.3$W$ and 1.54$W$) appear in the figure, for two pitch amplitudes (160 and 240 degrees), and two pitch frequencies (1.5 and 2.5 Hz). The figure shows how the $\varepsilon$ is mostly concentrated in the first POD mode of the stream-wise direction in all cases, with more than 70\%. In the cross-flow component, the energy is shared more uniformly mainly between the first three POD modes.

The decrease in thrust and propulsive velocity observed (Fig. \ref{Fpf}) for the $\theta_0=240^{\circ}$ case at frequencies higher than 2 Hz when the foil is away from the wall can also be explained through the POD analysis. Indeed, for frequencies higher than 2 Hz, since the momentum structure in the wake is then directed mostly perpendicularly to the swimming direction, the first POD mode at 2.5 Hz has dropped considerably if compared to the 1.5 Hz case (see second row and column in Fig. \ref{PODt}). Another point that can be seen is that, while for $\theta_0=160^{\circ}$  the energy of the different modes does not change noticeably when increasing the driving frequency, for $\theta_0=240^{\circ}$ on the contrary, the energy in the first POD mode does increase with frequency when the wall is present. The latter reflecting our previous observation that at these large angles the foil is diverting the momentum in a direction perpendicular to the propulsion direction and the presence of the wall reorients momentum favourably.

Figure \ref{PD} compares the first stream-wise POD mode (left column) and the average stream-wise velocity fields (right column) of the same cases. The trailing edge of the foil in its rest position is shown in the plots with a dashed white line for the sake of clarity. The POD and the averaged velocity fields appear normalised by the maximum value in the cases shown, for comparison. The plots certify again how the first stream-wise mode is enough to represent the momentum in the wake.

%**************************************************************************************************************
%**************************************************************************************************************
\section{Conclusions}

The experimental data presented in this work shows that swimming with large-amplitude undulatory motions at a moderate distance to a wall can have clear advantages in terms of velocity and thrust production. Positive ground (or wall) effect has been observed for the system presented here, when swimming with pitch motions of large amplitude ($\theta_0=160^{\circ}$ and $240^{\circ}$) and for distances to wall between 0.25 and 0.55 times the width ($W$) of the foil. Maximum improvements in velocity and thrust have been observed of about 25\% and 45\% respectively. The results also suggest that for distances of more than 1.5 widths of the foil, the ground effect can be neglected, fact also found by Blevins et al. \cite{Blevins:2013}. The fluid dynamical mechanisms behind this enhancement have been explored by investigating the flow field in the wake of foil, showing how the wall constrains the distribution of momentum in a direction favourable to propulsion. In addition to the analysis of the mean flow, which exhibits the constrained jet structure in the wake of the foil (Figs~\ref{avgC1} and \ref{avgC3}), the time-resolved vorticity fields show the changes in the wake vortex topology associated to the enhancement of propulsion ---e.g. Fig.~\ref{wk1} (a) and (c)---.

%In the final part of the paper we have explored the use of a POD technique to analyse the DPIV results. This analysis assessed the alterations in the distribution of $\varepsilon$ depending on the influence of the wall. Results of the POD confirmed the decrease in thrust away from the wall previously computed at the propulsive force and cruise velocity. Thereby, the use of POD method can be useful to study the changes in the distribution of flux.

As a point of perspective we can comment on the three-dimensional structure of the wake. Although the hypothesis of quasi-two-dimensionality underlying our analysis (as well as that of most of the literature on simplified model foils) can be partially justified alluding to the aspect ratio of the propulsive appendage, it is clear that the inherent 3D nature of this type of flows needs to be further analysed and included in realistic models. With respect to the present results, in addition to the vortex structures in the $xy$-plane analysed here, the wall will also affect the stream-wise structures in the $yz$-plane which have been recently established as important players in the drag-thrust balance \cite{Raspa:2014,Ehrenstein:2014}. These issues will be  the subject of future work.

The results with the present flexible foil excited by a pitching oscillation at its head are in agreement with what has been reported for a foil with heaving excitation \cite{Quinn:2014c}. This is an interesting observation from the point of view of bio-inspired design, where pitching motions associated to the elastic response of an appendage could sometimes be an optimal solution to actuate a robotic setup.

%**************************************************************************************************************

\section*{Appendix I}

A linear eigenvalue problem can be derived using the POD method. Let an ensemble of DPIV data \textbf{V}, with $N$ being the total number of the available flow fields or snapshots, arranged in column form i a way in which the first half of the columns are the stream-wise velocities and the second one the cross-flow velocities,

\begin{equation}\label{3a}
\mathbf{V}=\left [ \mathbf{v}^{1} \mathbf{v}^{2}...\mathbf{v}^{N} \right ]
\end{equation}

and the fluctuating part of the flow is,

\begin{equation}
\mathbf{V} = \mathbf{\tilde{V}}-{\bf \bar {v}} =\mathbf{\tilde{V}}-\frac{1}{N}\sum_{n=1}^N{\bf v}^n \qquad n=1,2, ... N
\label{E:fluctV}
\end{equation}

the eigenvalue formulation results in,

\begin{equation}
\mathbf{C}\mathbf{H}^i=\lambda^i \mathbf{H}^i
\label{E:EigSnpPOD}
\end{equation}

where the matrix $\mathbf{C}$ is,

\begin{equation}
\mathbf {C}=\mathbf{V}^T\mathbf{V}
\label{E:EigC}
\end{equation}

The solution of equation \ref{E:EigSnpPOD} consists of $N$ eigenvalues ($\lambda^i$) and the $N$x$N$ modal matrix ($\mathbf{H}$), made of column eigenvectors ($\mathbf{H}^n$). The eigenvectors provide a basis to produce the POD modes,

\begin{equation}
\phi^i=\frac{\sum_{n=1}^{N}H^i_n \mathbf{v}^n}{||\sum_{n=1}^{N}H^i_n \mathbf{v}^n||}, \qquad i=1,2,...N
\label{E:EigModes}
\end{equation}

where $||\cdot||$ denotes p2-norm, and it is calculated as the square root of the summation of the squares of each component inside the brackets. The result of equation \ref{E:EigModes} is a set of $N$ POD modes. As introduced in equation \ref{E:BasisPOD}, the flow can be expressed as a linear combination of POD modes and POD coefficients,

\begin{equation}
\mathbf{v}^n=\sum_{n=1}^N a_i^n \mathbf{\phi}^i=\mathbf{\Phi}\mathbf{a}^n
\label{E:FluctPartRec}
\end{equation}

hence, once the POD modes are available, the POD coefficients ($\mathbf{a}^n$) can be obtained,

\begin{equation}
\mathbf{a}^n=\mathbf{\Phi}^T\mathbf{v}^n
\label{E:PODcoef}
\end{equation}

This coefficients indicate how important is each POD mode in each time snapshot. The eigenvalues ($\lambda^i$), are proportional to the $\varepsilon$ of the fluctuating part of the flow and by sorting them in a decreasing fashion, $\lambda^i>\lambda^{i+1}$ for $i=1,...,N-1$ the most energetically important POD modes in the flow can be identified. The relative $\varepsilon$ associated to each POD mode can be calculated as,

\begin{equation}
\varepsilon_i=\frac{\lambda^i}{\sum_{n=1}^N \lambda^n}
\label{E:RelEner}
\end{equation}

%**************************************************************************************************************
\section*{References}

\bibliographystyle{unsrt}
%\bibliography{biblio_wall_effect}

\end{document}